\documentclass[aps,prl,amsmath,amssymb,amsfonts,showpacs,superscriptaddress,floatfix,twocolumn]{revtex4}

\usepackage{graphicx}
\usepackage{epsfig}
\usepackage{braket}
\usepackage{slashed}
\usepackage{bm}
\usepackage[usenames,dvipsnames]{color}

\newcommand{\kets}[1]{\ket{#1}_{n}}

\newcommand{\hstates}{\mathcal{H}}

\newcommand{\idm}{\openone}

\def\be{\begin{equation}}
\def\ee{\end{equation}}
\def\f#1#2{\frac{#1}{#2}}
\def\bea{\begin{eqnarray}}
\def\eea{\end{eqnarray}}

\begin{document}

\title{Matrix product states for gauge field theories}

\author{Boye Buyens}
\affiliation{Department of Physics and Astronomy,
Ghent
 University,
  Krijgslaan 281, S9, 9000 Gent, Belgium}

\author{Jutho Haegeman}
\affiliation{Department of Physics and Astronomy,
Ghent
 University,
  Krijgslaan 281, S9, 9000 Gent, Belgium}

  \author{Karel Van Acoleyen}
\affiliation{Department of Physics and Astronomy,
Ghent
 University,
  Krijgslaan 281, S9, 9000 Gent, Belgium}

\author{Henri Verschelde}
 \affiliation{Department of Physics and Astronomy,
Ghent
 University,
  Krijgslaan 281, S9, 9000 Gent, Belgium}

\author{Frank Verstraete}
  \affiliation{Department of Physics and Astronomy,
Ghent
 University,
  Krijgslaan 281, S9, 9000 Gent, Belgium}
  \affiliation{Vienna Center for Quantum Science and Technology, Faculty of Physics, University of Vienna, Boltzmanngasse 5, 1090 Vienna, Austria}

\begin{abstract}
The matrix product state formalism is used to simulate Hamiltonian lattice gauge theories. To this end, we define matrix product state manifolds which are manifestly gauge invariant. As an application, we study 1+1 dimensional one flavor quantum electrodynamics, also known as the massive Schwinger model, and are able to determine very accurately the ground state properties and elementary one-particle excitations in the continuum limit. In particular, a novel particle excitation in the form of a heavy vector boson is uncovered, compatible with the strong coupling expansion in the continuum. We also study full quantum non-equilibrium dynamics by simulating the real-time evolution of the system induced by a quench in the form of a uniform background electric field.

\end{abstract}

\maketitle

Gauge theories hold a most prominent place in physics. They appear as effective low energy descriptions at different instances in condensed matter physics and nuclear physics. But far and foremost they lie at the root of our understanding of the four fundamental interactions that are each mediated by the gauge fields corresponding to a particular gauge symmetry. At the perturbative quantum level, this picture translates to the Feynman diagrammatic approach that has produced physical predictions with unlevelled precision, most famously in quantum electrodynamics (QED). However the perturbative approach miserably fails once the interactions become strong. This problem is most pressing for quantum chromodynamics (QCD), where all low energy features like quark confinement, chiral symmetry breaking and mass generation are essentially non-perturbative.

Lattice QCD, which is based on Monte Carlo sampling of Wilson's Euclidean lattice version of gauge theories,
has historically been by far the most successful method in tackling this strongly coupled regime. Using up a sizable fraction of the global supercomputer time, state of the art calculations have now reached impressive accuracy, for instance in the ab initio determination of the light hadron masses \cite{Durr}. But in spite of its clear superiority, the lattice Monte Carlo sampling also suffers from a few drawbacks. There is the infamous sign problem that prevents application to systems with large fermionic densities. In addition, the use of Euclidean time, as opposed to real time, presents a serious barrier for the understanding of dynamical non-equilibrium phenomena. Over the last few years there has been a growing experimental and theoretical interest in precisely these elusive regimes, e.g. in the study of heavy ion collisions or early time cosmology.

In this letter we study the application of tensor network states (TNS) as a possible complementary approach to the numerical simulation of gauge theories. This is highly relevant as this Hamiltonian method is free from the sign problem and allows for real-time dynamics. As a first application we focus on the massive Schwinger model. For this model the TNS approach has been studied before by Byrnes et al \cite{Byrnes} and Ba\~{nuls} et al \cite{Banuls}. By integrating out the gauge field (which one can only do for $d$=1+1), the model was reduced to an ordinary spin model, yet with a non-local Hamiltonian. Our approach is conceptually different, as we keep the gauge field degrees of freedom, which enables us to take the thermodynamic limit, with the relevant global symmetries exact. TNS have been considered also for the discrete $Z2$ gauge theory, for $d$=1+1 by Sugihara \cite{Sugihara}, and for $d$=2+1 by Tagliacozzo and Vidal \cite{Tagliacozzo}.

Over the last decade the TNS framework has emerged as a powerful tool for the study of local quantum many body systems, exploring the fact that physical states (i.e. ground states and their low energy excitations) only occupy a tiny corner of the full Hilbert space \cite{Cirac}. This is exemplified by the relatively small amount of quantum entanglement that these states possess. TNS are then trial quantum states that precisely capture this feature, allowing for relatively low cost numerical variational calculations. In one spatial dimension, they also go by the name of matrix product states (MPS), underlying the well known density matrix renormalization group algorithm (DMRG) \cite{White}. At present MPS/DMRG is the state of the art
 method in the numerical study of both static and dynamical properties of $d$=1+1 strongly correlated condensed matter systems. And also in higher dimensions the TNS framework \cite{Verstraete}, although less developed, is considered to be a promising candidate for the numerical simulation of strongly interacting quantum many body sytems.

The essential new ingredient with respect to the usual MPS applications on quantum many body systems is that for the Hamiltonian formulation of gauge theories, of the full Hilbert space only the subspace of gauge invariant states is actually physical. Although, due to Elitzur's theorem \cite{Elitzur} gauge invariance will not be broken on the full Hilbert space, there will be typically many more low-energy excitations in the full space than in the constrained physical subspace. It is therefore crucial to restrict the variational MPS manifold to this physical subspace. Notice that the very same issue poses itself in the context of the simulation of gauge theories with ultracold atoms. See \cite{Zohar} for a recent proposal to implement gauge invariance in that case.

The massive Schwinger model is QED in 1+1 dimensions, with one flavor of fermionic particles with mass $m$, interacting through a $U(1)$ gauge field with coupling $g$ (which has mass dimension one for $d$=1+1). This model shares some interesting features with QCD, most notably the fermions are confined into zero charge bound states. Furthermore in the continuum it can be studied by a strong coupling expansion \cite{Coleman, Adam}, which makes it a perfect benchmark model.  We will apply our gauge invariant MPS construction on the Hamiltonian lattice formulation of the model, focusing on the strongly coupled regime $g/m \gtrsim 1$, and extrapolating our results to the continuum.  We determine the ground state and stable bound states. In addition, we show how our formalism indeed allows for the study of real time phenomena and simulate the full quantum dynamics induced by a background electric field.

\noindent \emph{The Schwinger Hamiltonian}
\noindent To write down a lattice Hamiltonian for the Schwinger model, one starts from the Lagrangian density in the continuum: \be \mathcal{L} = \bar{\psi}\left(\gamma^\mu(i\partial_\mu+g A_\mu) - m\right) \psi - \frac{1}{4}F_{\mu\nu}F^{\mu\nu}\,. \ee One can then perform a Hamiltonian quantization in the time-like axial gauge ($A_0=0$), which can be turned in a lattice system by the Kogut-Susskind spatial discretization \cite{Kogut} with the two-component fermions sited on a staggered lattice. These fermionic degrees of freedom can then finally be converted to spin 1/2 degrees of freedom by a Jordan-Wigner transformation, leading to the gauged spin Hamiltonian (see \cite{Byrnes} for more details): \bea\label{equationH} H&=& \frac{g}{2\sqrt{x}}\Biggl( \sum_{n \in \mathbb{Z}} {L}(n)^2 + \frac{\mu}{2} \sum_{n \in \mathbb{Z}}(-1)^n(\sigma_z(n) + (-1)^{n}) \nonumber
\\ &&+ x \sum_{n \in \mathbb{Z}}(\sigma^+ (n)e^{i\theta(n)}\sigma^-(n + 1) + h.c.)\biggl).\eea Here we have introduced the parameters $x \equiv 1/(g^2a^2)$  and $\mu \equiv 2\sqrt{x}m/g$, with $a$ the lattice spacing.

The spins live on the sites of the lattice, with $\sigma_z(n)\ket{s_n}= s_n \ket{s_n} (s_n=
\pm 1)$, and $\sigma^{\pm}=1/2(\sigma_x\pm i \sigma_y)$ the spin ladder operators. Notice the different second (mass) term in the Hamiltonian for even and odd sites. This can be traced back to the staggered formulation, with the even sites being reserved for the `positrons' and the odd sites for the `electrons'. For the even positron sites $s_{2n}=+1$ can be viewed as an occupied state, while $s_{2n}=-1$ corresponds to an empty state, and vice versa for the odd electron sites. The gauge fields $\theta(n)=a g A_1(na/2)$, live on the links between the sites. Their conjugate momenta $L(n)$, with $[\theta(n),L(n')]=i\delta_{n,n'}$, correspond to the electric field, $gL(n)=E(na/2)$. Since $\theta(n)$ is an angular variable, $L(n)$ will have integer charge eigenvalues $p_n \in \mathbb{Z} $. The local Hilbert space, spanned by the corresponding eigenkets $\ket{p_n}$ is therefore infinite, but in practice we will do a truncation and consider $|p_
 n|\leq p
 _{max}$ in a numerical scheme. For our calculations we take $p_{max}=3$.

The Hamiltonian (\ref{equationH}) is invariant under $T^2$, a translation over two sites, and the corresponding eigenvalues read $T^2=e^{2 i k a}$, where $k$ is the physical momentum of the state. Another symmetry that will be useful is $CT$, obtained by a translation over one site, followed by a charge conjugation, $C\ket{s_n,p_n}=\ket{-s_n,-p_n}$.
Since $C^2=1$, we will have $CT=\pm e^{ika}$. The states with positive sign then correspond to the scalar sector, while the negative sign corresponds to the vector sector.

In addition, the Hamiltonian is invariant under the residual time-independent local gauge transformations, generated by  \begin{equation}
 G_n=L(n)-L(n-1)-\frac{1}{2}( \sigma_z(n) + (-1)^n )\,.
\label{spingauss}
\end{equation}
It is this gauge invariance that sets the Hamiltonian quantization of gauge theories apart from the Hamiltonian quantization of ordinary systems.  For gauge theories only the subspace of gauge invariant states will be physical: $G_n\ket{\Psi}_{phys}=0$ for every $n$. This is called the Gauss' law constraint, as $G_n=0$ is indeed the discretized version of $\partial_x E=\rho$. We will now show how one can tailor the MPS formalism towards a constrained variational method on this physical gauge invariant subspace.

\noindent\emph{Gauge invariant MPS.}
A general, not necessarily gauge invariant MPS for the lattice spin-gauge system (\ref{equationH}) has the form:  \be \sum_{s_n,p_n}Tr[B_1^{s_1}C_1^{p_1} B_2^{s_2}C_2^{p_2}\ldots C_{2N}^{p_{2N}}] \ket{s_1,p_1,s_2,p_2\ldots, p_{2N}}\,,\label{MPS}\ee
where for now we consider a finite lattice of $2N$ sites. Here, each $B_n^{s_n}$ (and $C_n^{p_n}$) is a complex $D\times D$ matrix with components $[B_n^{s_n}]_{\alpha \beta}$, that constitute the variational parameters of the trial state. The indices $\alpha,\beta=1,\ldots D$ are referred to as virtual indices, and $D$ is called the bond dimension.

Gauss' law (see (\ref{spingauss})) prescribes how to update the electric field $L(n)$ at the right link of a site $n$: either staying with the value $L(n-1)$ at the left in case there is no charge at the site, or adding/subtracting one unit in case there is a positive/negative charge at the site. This can be conveyed by the matrix multiplications in an MPS by giving the virtual indices a multiple index structure $\alpha\rightarrow (q,\alpha_q)$, where $q$ labels the charge, and taking the matrices of the form:
\bea
{[B_{n}^{s_n}]}_{(q,\alpha_q),(r,\beta_r)} &=& {[b_{n,q}^{s_n}]}_{\alpha_q,\beta_r}\delta_{q+(s_n+(-1)^n)/2,r}\nonumber\\
{[C_{n}^{p_n}]}_{(q,\alpha_q),(r,\beta_r)} &=& [c_n^{p_n}]_{\alpha_q,\beta_r}\delta_{q,p_n}\delta_{r,p_n}
\label{gaugeMPS}\,.\eea
 One can readily verify that an MPS (\ref{MPS}) with matrices of this form, indeed obeys the Gauss' law constraint at every site. Conversely, we show in S1 \cite{SM} that every gauge invariant state $\ket{\Psi}$, obeying $G_n\ket{\Psi}=0$ for every $n$, has an MPS representation of the form (\ref{gaugeMPS}).

\begin{figure}
\begin{tabular}{rr}
  \includegraphics[width=42mm]{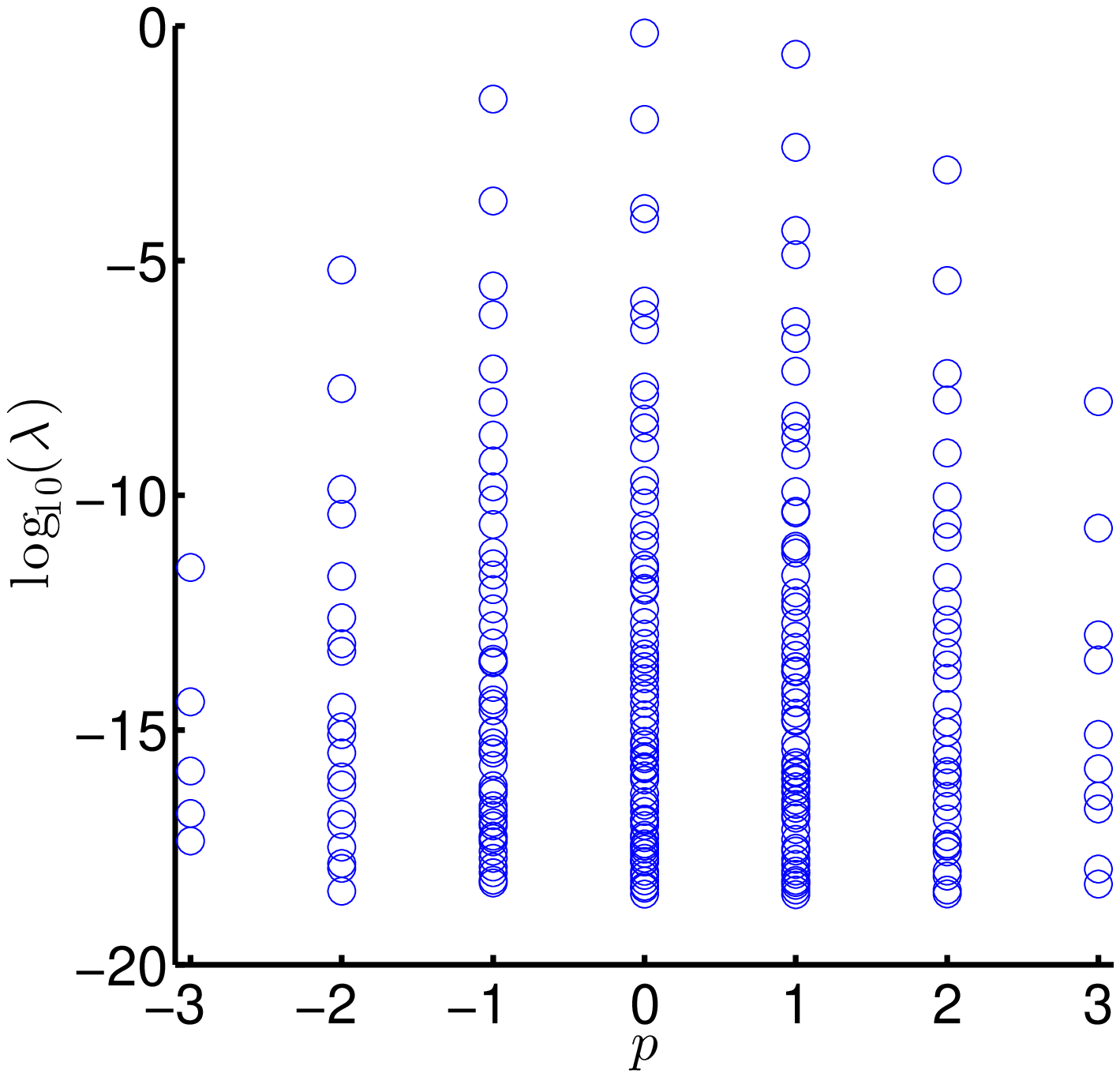} &   \includegraphics[width=43mm]{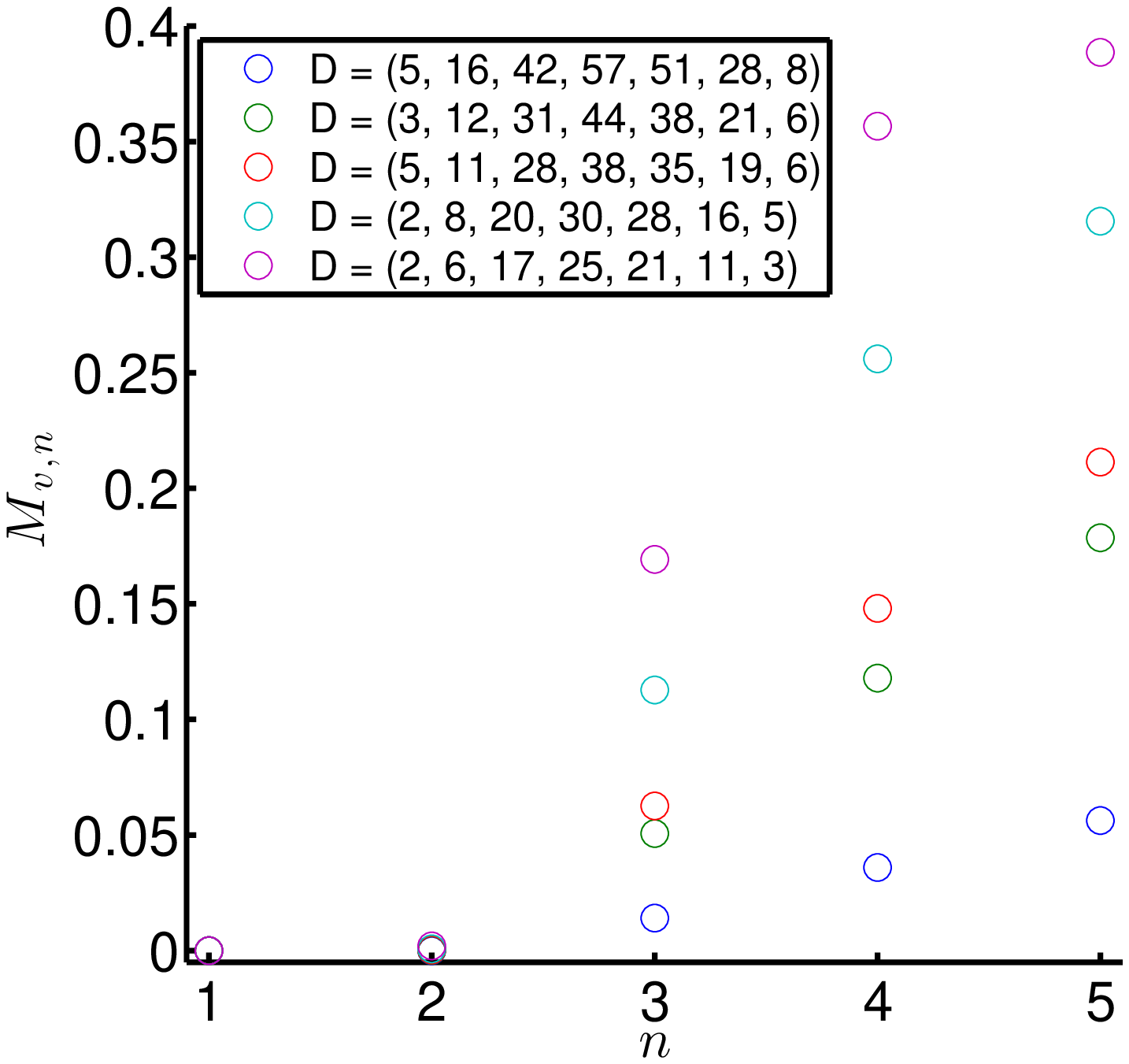}
\end{tabular}
\caption{\label{fig:Res1}Results for $m/g = 0.25$, $x = 100$. Left (a): distribution of the (base-10) logarithm of the Schmidt coefficients $\lambda$ in every charge sector for $D = (5, 20, 48, 70, 62, 34, 10)$. Right (b): Difference for the estimated energies of the excited states for various bond dimension with respect to these with $D = (5, 20, 48, 70, 62, 34, 10)$  for the vector sector $\gamma = -1$. Only the first two excitations are stable under variation over $D$.   }
\end{figure}

\noindent \emph{Ground state and excitations.} To obtain a ground-state approximation in the thermodynamic limit ($N\rightarrow \infty$) it will be useful to block a site and link into one site with local Hilbert spaced spanned by the states $\ket{q_{2n-1}} = \ket{s_{2n-1}, p_{2n-1}}$ and $\ket{q_{2n}} = \ket{s_{2n}, p_{2n}}$. Anticipating $CT = 1$ the gauge-invariant  ground state ansatz then takes a form similar to a uniform MPS (uMPS)\cite{Nachtergaele}:
\begin{equation}\ket{\Psi(A)} = \sum_{q_n} \bm{v}_L^\dagger \left(\prod_{n \in \mathbb{Z}}A^{q_n}\right) \bm{v}_R \ket{\bm{q}^c}, \end{equation} where $\ket{\bm{q}^c} = \ket{\{(-1)^n q_{n}\}_{n \in \mathbb{Z}}}$, $v_L, v_R \in \mathbb{C}^D$, and $A^{q} \in \mathbb{C}^{D \times D}$ as follows from (\ref{gaugeMPS}):
\be [A^{s,p}]_{(q,\alpha_q);(r,\beta_r)} =  [a^{s,p}]_{\alpha_q, \beta_r}\delta_{p, q + (s+1)/2} \delta_{r,-p}\,. \label{GIA}\ee
We refer to S2 \cite{SM} for the details and the implementation of the time-dependent variational principle (TDVP) \cite{HaegemanTDVP} to obtain an approximation for the ground state.
The variational freedom of the gauge invariant state $\ket{\Psi(A)}$ lies within the matrices $a^{s,p} \in \mathbb{C}^{D_q \times D_r}$ and the total bond dimension of the uMPS equals $D = \sum_{q \in \mathbb{Z}}D_q$. It will be important to choose the distribution of $D_q$ wisely, according to the relative weight of the different charge sectors. As illustrated in fig.1a,  this is done by looking at the Schmidt coefficients for an arbitrary cut, and demanding that the smallest coefficients of each sector coincide more or less. The resulting distribution of $D_q$ is peaked around $q=0$, and justifies our $p_{max}=3$ truncation that corresponds to $D_q=0$ for $|q|>3$ (see also S3 \cite{SM}).

Once we have a good approximation for the ground state, we can use the method of \cite{HaegemanEA, HaegemanBMPS} to obtain the one-particle excited states.  The excitations are labelled by their (physical) momentum $k \in [-\pi/2a, \pi/2a[$ and their $CT$ quantum number $\gamma=\pm 1$.  For a given ground-state approximation we then take the following ansatz state $\ket{\Phi_{k,\gamma}(B,A)}$ for the one-particle excitations:
\be\sum_{m\in\mathbb{Z}}\hspace{-0.1cm}e^{ikma}\gamma^{m}\hspace{-0.1cm}\sum_{q_n}\hspace{-0.1cm}\bm{v}_L^\dagger\!\! \left(\prod_{n < m}A^{q_n}\right)\!\!B^{q_m}\!\!\left(\prod_{n > m}A^{q_n}\right)\!\! \bm{v}_R\ket{\bm{q}^c }\,, \ee
with $B^q$ again of the gauge invariant form (\ref{GIA}) with general matrices $b^{s,p}$. These are determined variationally by
minimizing their energy in the ansatz subspace which leads to a generalized eigenvalue problem (see S2 \cite{SM} for more details). For a given momentum and $CT$ quantum number we typically find different local minima of which only one or two are stable under variation of the bond dimension $D$ (see fig.1b). It are these stable states that we can interpret as approximations to actual physical one-particle excitations.

In table 1 we display our results for the continuum extrapolations $a\rightarrow0$ ($x\rightarrow \infty$) of the ground state energy density and the mass of the different one-particle excitations (we refer to S3 \cite{SM} for more details). For $g/m\neq 0$ we find three excited states, one scalar and two vectors, with the hierarchy of masses $M_{v,1}<M_{s,1}<M_{v,2}$ matching that of the strong coupling result \cite{Coleman,Adam}. This is the first time that the second vector excitation has been found numerically. For the energy density and the two lowest mass excitations our results are consistent with the previous most precise simulations \cite{Byrnes,Banuls}, with a similar or sometimes better accuracy. As shown in the supplementary material, we were also able to reconstruct the Einstein dispersion relation for small momenta $ka\ll 1$.

\begin{table}
\begin{tabular}{|c||c|c|c|c|c|}
\hline
$m/g$ & $\omega_0$ & $M_{v,1}$ & $M_{s,1}$ & $M_{v,2}$   \\
\hline
0 &-0.318320(4)         &  0.56418(2)&   &  \\
0.125&        -0.318319(4) &0.789491(8)  &1.472(4)  &2.10 (2)  \\

0.25 &-0.318316(3)         &  1.01917 (2)&  1.7282(4)&2.339(3)  \\
0.5 &        -0.318305(2) &  1.487473(7)& 2.2004 (1) & 2.778 (2) \\
0.75 &        -0.318285(9) &  1.96347(3)& 2.658943(6) &3.2043(2)  \\
1 &-0.31826(2)&2.44441(1)         &3.1182 (1)  & 3.640(4)  \\
\hline

\end{tabular}
\caption{Energy density and masses of the one-particle excitations (in units $g=1$) for different $m/g$. The last column displays the result for the heavy vector boson, compatible with the prediction of Coleman \cite{Coleman,Adam}}
\end{table}

\noindent \emph{Real-time evolution.} One of the main advantages of the TNS framework is that it allows for the full quantum simulation of real-time phenomena. Specifically we have investigated the non-equilibrium dynamics induced by applying a uniform electric field $E$ on the ground-state $\ket{\Psi_0}$ at time $t=0$.
Physically, the process corresponds to the so called Schwinger particle creation mechanism \cite{Schwinger2}, but now for a confining theory. This process has been studied extensively in the past, either with some effective classical kinetic description \cite{classical, Kluger}, or in the semi-classical limit for the gauge fields \cite{Kluger,Hebenstreit,Gelis} and recently also with the AdS/CFT correspondence \cite{Kawai}. Here we focus on the systematics of our method and present some first results, allowing us to validate our formalism against the predicted scaling from linear response theory and energy conservation. A more detailed analysis will be presented elsewhere \cite{Buyens}.

In our set-up the application of a uniform electric field is simulated by applying a uniform quench, replacing $L(n)$ with $L(n)+\alpha$ in the Hamiltonian (\ref{equationH}), where $E=g\alpha$. As before we define our ansatz by blocking a site and a link into one site. But since the background field now breaks $CT$ invariance we can only anticipate translation symmetry $T^2=1$ over two sites. Our ansatz thus takes the form: \be \ket{\Psi(A_1,A_2)} = \sum_{q_n} \bm{v}_L^\dagger (\prod_{n \in \mathbb{Z}}A_1^{q_{2n-1}}A_2^{q_{2n}}) \bm{v}_R \ket{\{{(q_{2n-1},q_{2n})\}}} \ee where $q_n = (s_n,p_n)$. From (\ref{gaugeMPS}) it follows that gauge invariance is imposed if we set $[A_n^{s,p}]_{(q,\alpha_q);(r,\beta_r)} =  [a_n^{s,p}]_{\alpha_q, \beta_r}\delta_{p, q + (s+(-1)^n)/2} \delta_{r,p}$.

\begin{figure}
\begin{tabular}{rr}
 \includegraphics[width=43mm]{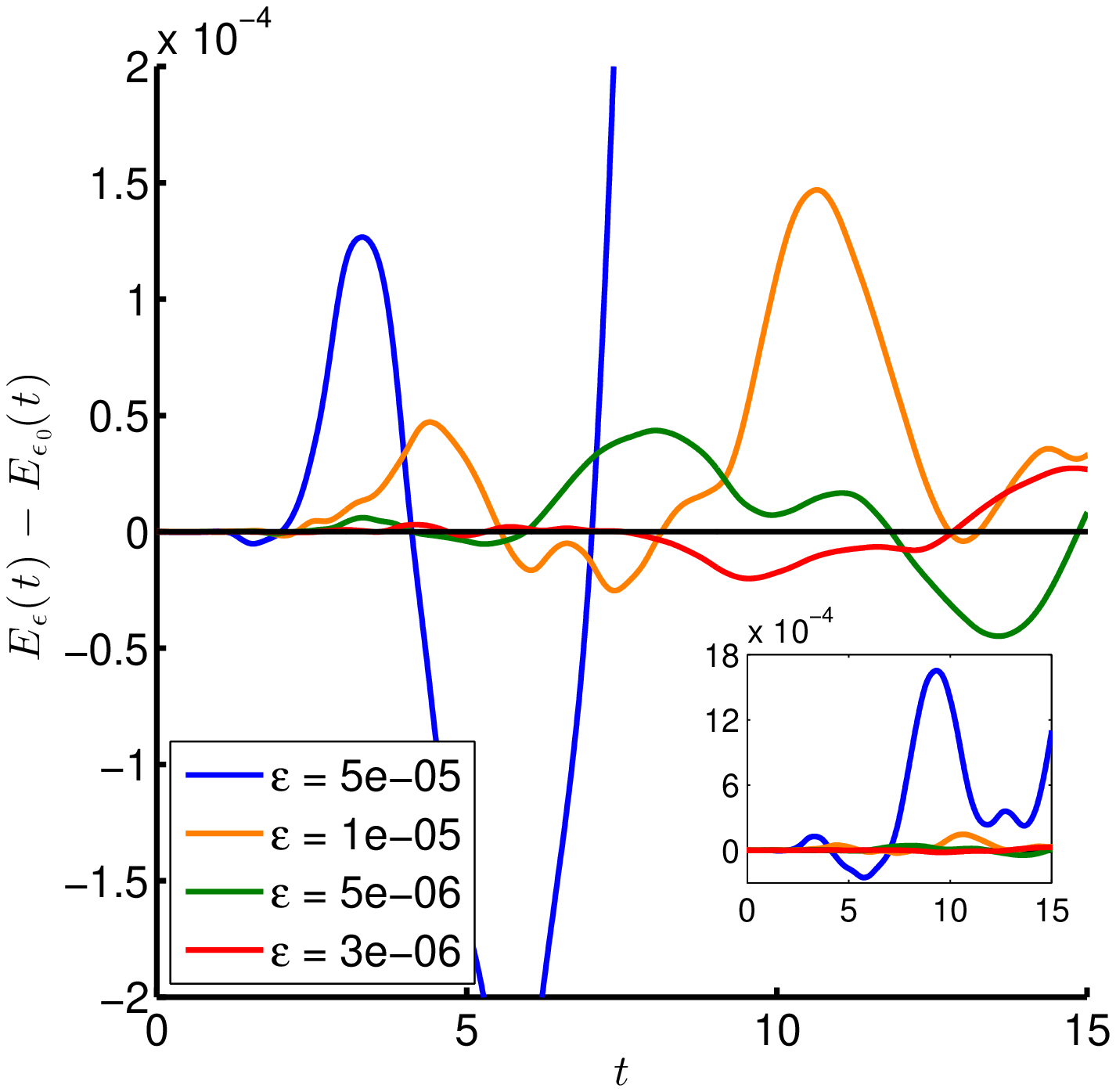} &   \includegraphics[width=43mm]{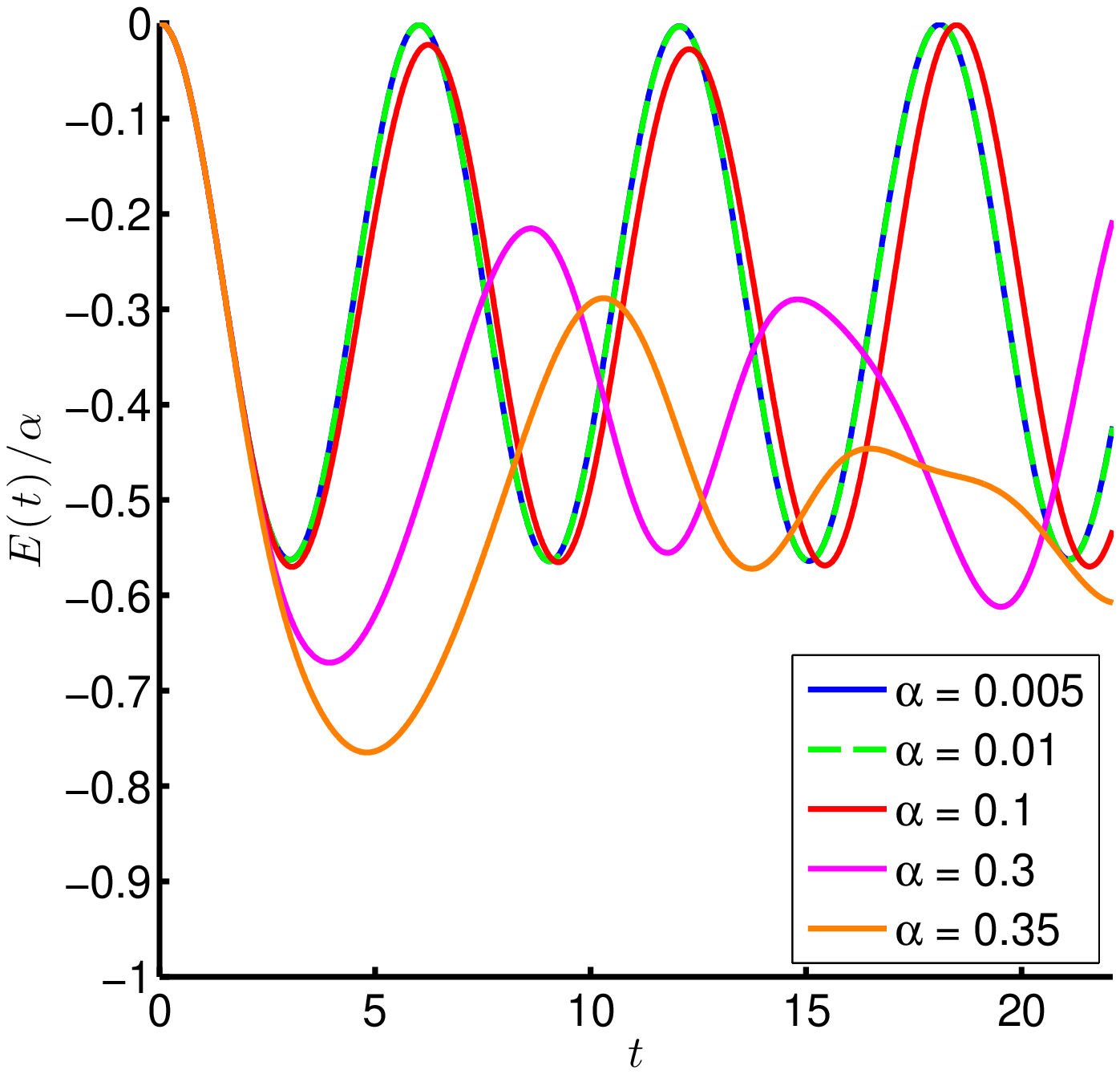}\\
  \includegraphics[width=43mm]{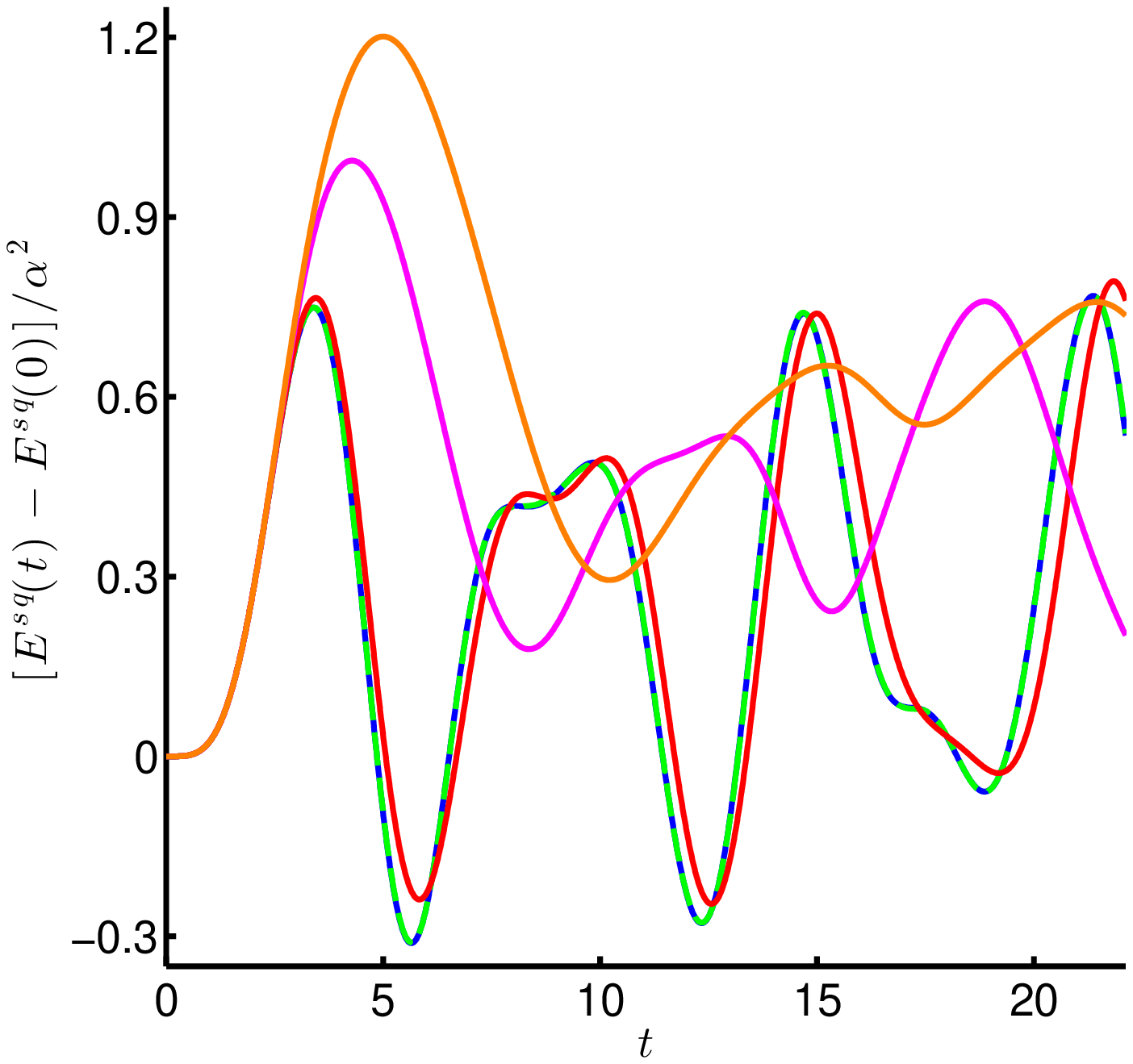} &   \includegraphics[width=43mm]{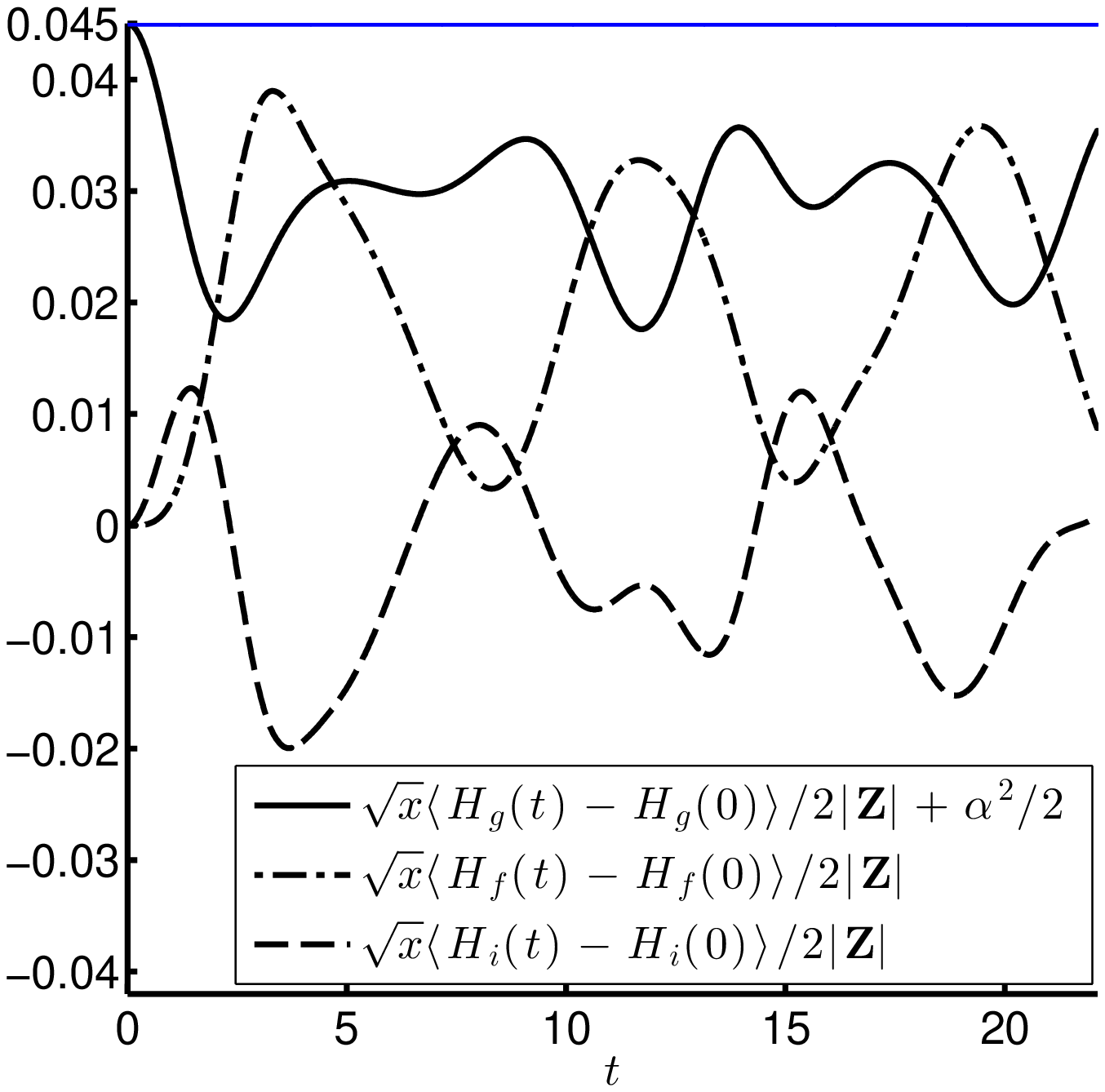}\\
\end{tabular}
\caption{Results for $m/g= 0.25, x = 100$, all quantities are in units $g=1$. Left up (a): Difference of $E(t)\equiv\bra{\Psi_0(t)}L(1)+L(2) \ket{\Psi_0(t)}/2$ for various tolerances $\epsilon$ with respect to the estimated value for $\epsilon =\epsilon_0= 2\cdot 10^{-6}$ ($\alpha = 0.3$). Right up (b): $E(t)/\alpha$ for different values of $\alpha$. Down left (c): $E^{sq}(t)/\alpha^2$, with $E^{sq}(t)\equiv \bra{\Psi_0(t)}L^2(1)+L^2(2) \ket{\Psi_0(t)}/2 $, for the same set of $\alpha$-values as in (b). Down right (d): for $\alpha=0.3$, the different energy densities of resp. the first ($H_g$), second ($H_f$) and third term ($H_i$) of (\ref{equationH}) but with $L(n) \rightarrow L(n) + \alpha$ (we subtracted the values at $t = 0$ without background field.). The straight blue line is the total energy density obtained as the sum of the three terms.    }

\end{figure}

To perfom the real-time evolution we have implemented the infinite time-evolving block decimation algorithm (iTEBD) \cite{iTEBD} using a fourth-order Trotter expansion \cite{TrotterSuzuki} with time-step $  dt = 0.01/g$.  We refer to S4 \cite{SM} for the details. At every step iTEBD truncates the Hilbert space by discarding the Schmidt coefficients lower than some fixed threshold $\epsilon^2$. For gauge invariant MPS this in turn determines the required bond dimensions $D_p$ for every charge sector, that will evolve in time. For instance, for the value $\epsilon_0=2\cdot 10^{-6}$ that we used for the simulations in figs. 2b-2d, and for $\alpha=0.3$, the maximal bond dimension goes from $D_0=18$ at $t=0$ to $D_0=173$ at $t=25$. It is this growth of the required bond dimensions, which can be traced back to the growth of entanglement, that makes the computations more costly at later times. As the simulation should be exact as $\epsilon \rightarrow 0$, the convergence in $\epsilon$ can be used to control the truncation error for a certain observable. We illustrate this in fig. 2a for the electric field expectation value. Also notice that the convergence rate decreases in time. Keeping the truncation error small for larger time intervals will therefore require smaller values of the tolerance $\epsilon$.

In fig. 2b we display our result for the evolution of the electric field expectation value (minus the background value) for different values of $\alpha$. For early times we clearly find the $\alpha$-scaling behavior as predicted from linear response theory (see S5 \cite{SM}). The $\alpha = 0.005$ and $\alpha = 0.01$ cases remain in the linear response regime throughout the entire depicted evolution; the periodic oscillations in this case can be traced back to the dominant production of the single-particle vector excitation in the linear response regime (see S5). Larger values of $\alpha$ progressively depart from linear response, showing more complex behavior at later times. The physical picture here \cite{Kluger}, that we study in more detail in \cite{Buyens}, is that the charged fermionic particles that are created by a strong initial electric field, in turn backreact onto this field.

In fig. 2c we display the analogous result for the electric field squared expectation value. As the operator $\sum_n L^2(n)$ is now invariant under $CT$, this should scale as $\alpha^2$ for early times (see S5), which is indeed what we find. Finally, in fig. 2d we show the evolution of the energies in the different sectors. We see that the energy which is initially injected in the first gauge field term in (\ref{equationH}), partially leaks into the second fermionic mass term and third kinetic/interaction term, as we can again qualitatively understand from the fermionic particle creation picture. In \cite{Hebenstreit} a similar behavior was observed in the semi-classical limit. A last cross-check of our real-time results is then provided by the total energy conservation which is indeed satisfied as can be seen from the blue line in fig. 2d.

\noindent \emph{Conclusions.} In this letter we have demonstrated the potential of MPS as numerical method for gauge theories. It is clear that we have only scratched the surface of this approach and that even within the Schwinger model there are many other types of calculations one could do, like for instance the construction of two-particle scattering states \cite{Laurens}. Looking further afield, one can easily generalize our gauge invariant MPS ansatz to other gauge groups like $SU(N)$ and also to higher dimensions. Explicitly for $d$=2+1, the gauge invariant 2$d$ PEPS \cite{Verstraete} construction now involves five-leg tensors with four virtual indices and one physical index ($c$ = charge) on the sites, of the form ${[B^{c}]}_{(q_l,\alpha_{q_l}),(q_r,\alpha_{q_r}),(q_d,\alpha_{q_d}),(q_u,\alpha_{q_u})} = {[b_{q_l,q_r,q_u}^{c}]}_{\alpha_{q_l},\alpha_{q_r},\alpha_{q_d},\alpha_{q_u}}\delta_{q_l+q_d+c,q_r+q_u}$, while on the
 links we
  get a three-leg tensor with two virtual indices and one physical index ($p$ = electric field unit) ${[C^{p}]}_{(q_l,\alpha_{q_l}),(q,\alpha_{q_r})} = [c^{p}]_{\alpha_{q_l},\alpha_{q_r}}\delta_{q_l,p}\delta_{q_r,p}$.

While preparing our manuscript the paper \cite{Rico} appeared with an approach that is conceptually close to ours.  There the authors use a quantum link model to write down gauge invariant MPS for the Schwinger model.

\noindent\emph{Acknowledgements.}
We thank Mari-Carmen Ba\~{n}uls  for suggesting us to look at real-time quench dynamics. Furthermore we acknowledge very interesting discussions with Mari-Carmen Ba\~{n}uls, David Dudal and Lucca Tagliacozzo. This work is supported by an Odysseus grant from the FWO, a PhD-grant from the FWO (B.B), the FWF grants FoQuS and Vicom, the ERC grant QUERG and the EU grant SIQS.

\newpage

\onecolumngrid

\onecolumngrid
\appendix

\section{Supplemental Material}
\section{S1: Gauge invariant states}\label{app1}
\noindent Consider a lattice with $2N$ sites. The basis of the total Hilbert space $\hstates$ is $\{ \ket{\bm{q}} \equiv \ket{\{s_n, p_n\}_{1 \leq n \leq 2N}} :s_n = \pm 1, p_n \in \mathbb{Z} \}$. A general state $\ket{\Psi} \in \hstates$ can be written as a MPS in the canonical form (see \cite{PerezGarcia}, theorem 1):
\begin{equation}\label{MPSSw1}\ket{\Psi} = \sum_{\{s \} = \pm 1 }\sum_{\{p \} \in \mathbb{Z} } \prod_{n = 1}^{N}B^{s_{2n - 1}}_{2n - 1}C^{p_{2n - 1}}_{2n- 1}B^{s_{2n}}_{2n}C^{p_{2n}}_{2n}\ket{\bm{q}},\end{equation}
where $B_k^{s_k} \in \mathbb{C}^{D^k  \times D'^{k}}$, $C_k^{p_k} \in \mathbb{C}^{D'^{k} \times D^{k+1}}$ and $D^1  = D^{2N+1} = 1$. By `being in its canonical form' we mean that
\begin{equation}\label{Scanform1}\sum_{s = \pm 1}  B^{s}_n {B^{s}_n}^\dagger = \idm,  \sum_{p \in \mathbb{Z}}  C^{p}_n {C^{p}_n}^\dagger = \idm\end{equation}
and that there exist positive definite diagonal matrices $l^C_n$ and $l^B_n$ such that
\begin{equation}\label{Scanform2}\sum_{s= \pm 1}  {B^{s}_n}^\dagger
l^B_n B^{s}_n  = l^C_n,  \sum_{p \in \mathbb{Z}}  {C^{p}_n}^\dagger l^C_{n} C^{p}_n  = l^B_{n+1}.\end{equation}
Because QED is a gauge theory we have to restrict to $\hstates_{phys}$, the set of all gauge invariant states. This means that every $\ket{\Psi} \in \hstates_{phys}$ has to satisfy
\begin{equation}\label{Gausslawbis}\ket{\Psi} =  \exp(-i\varphi_lG_l)\ket{\Psi}, l = 1,\ldots,2N.\end{equation}
where
\begin{equation}G_l = L(l) - L(l-1) - \frac{\sigma_z(l) + (-1)^l}{2}, L(0) = 0. \end{equation}
The right-hand side can also be written as a MPS with the same bond dimension:
\begin{equation} \exp(-i\varphi_lG_l)\ket{\Psi} =\sum_{\{s \} = \pm 1 }\sum_{\{p \} \in \mathbb{Z} }  \mbox{tr}\Biggl( \Biggl[\prod_{n = 1}^{N}\tilde{B}^{s_{2n - 1}}_{2n - 1}\tilde{C}^{p_{2n - 1}}_{2n- 1}\tilde{B}^{s_{2n}}_{2n}\tilde{C}^{p_{2n}}_{2n}\Biggl] \Biggl)\ket{\bm{q}}\end{equation}
where if $l > 1$: $\tilde{B}_k^{s_k} = B_k^{s_k}$  for $k \neq l $, $\tilde{C}_k^{s_k} = C_k^{s_k}$ for $k \neq l-1, l$ and $\tilde{C}_{l-1}^{p} = e^{-i\varphi_{l}p}C_{l-1}^{p}, \tilde{B}_{l}^{s} = e^{-i\varphi_l[\frac{s + (-1)^l}{2}]}B_l^{s},\tilde{C}_{l}^{p} = e^{i\varphi_{l}p}C_l^{p}$ and if $l = 1$: $\tilde{B}_k^{s_k} = B_k^{s_k}$, $\tilde{C}_k^{s_k} = C_k^{s_k}$ for $k \neq 1$ and $\tilde{B}_{1}^{s} = e^{-i\varphi_1[\frac{s -1}{2}]}B_1^{s},\tilde{C}_{1}^{p} = e^{i\varphi_{1}p}C_1^{p}$. Because the two MPS with identical bond dimension have to represent the same state $\ket{\Psi}$ and (\ref{MPSSw1}) is assumed to be in the canonical form, it follows by \cite{PerezGarcia}, theorem 2, that there exists invertible square matrices $U_n$ and $V_n$ such that $\tilde{B}_n^{s_n} = U_n^{-1}B_n^{s_n}V_n$, $\tilde{C}_n^{s_n} = V_n^{-1}C_n^{s_n}U_{n+1}$ where $U_1 = 1$ and $U_{2N + 1} = 1$. Note that $U_n$ and $V_n$ depend on $\varphi_l$ and $l$.

The matrices $U_n$ and $V_n$ are unitary matrices. Indeed, it's not hard to check that $\tilde{B}_n^{s}$ and $\tilde{C}_n^{p}$ also obey (\ref{Scanform1}) and (\ref{Scanform2}). For $n = 2N$ we have that $\tilde{C}_{2N}^{s} = V_{2N}C_{2N}^s$ which implies that $V_{2N} V_{2N}^\dagger = \idm$. Using this, $\tilde{B}_{2N}^p = U_{2N}^{-1}B_{2N}^pV_{2N}^p$ and the fact that $B_{2N}^p$ and $\tilde{B}_{2N}^p$ obey (\ref{Scanform1}) it follows that $U_{2N}U_{2N}^\dagger = \idm$. Proceeding in the same way from $n = 2N$ till $n = 1$ one sees that all the matrices $U_n$ and $V_n$ are unitary.

Now we will prove that $U_n, V_n = \idm$ for $n \neq l$. If $n < l$ we may assume that $l > 1$. Note that $U_1 = 1$ and that $B_1^s = \tilde{B}_1^s = B_1^sV_1$. Using (\ref{Scanform2}) it follows that $V_1 = (l_1^C)^{-1}\sum_{s}{B_1^s}^\dagger l_1^B B_1^s = \idm.$ Assume now $V_{n-1} = \idm$ ($n < l-1$) then $C_{n-1}^p = \tilde{C}_{n-1}^p = C_{n-1}^pU_{n}$, which implies $U_n = (l_n^B)^{-1} \sum_p {C_{n-1}^p}^\dagger l_{n-1}^C C_{n-1}^p = \idm$, i.e. $V_{n-1} = \idm$ implies that $U_n = \idm$ for $n < l - 1$. Using the same ideas one proofs that $U_n = \idm$ implies $V_{n} = \idm$ ($n < l $). This concludes the case $n < l$. For $n > l$ one starts from $\tilde{C}_{2N}^p = C_{2N}^p = V_{2N}C_{2N}^p$. From (\ref{Scanform1}), we obtain that $V_{2N} = \idm$. As a consequence $\tilde{B}_{2N-1}^s = B_{2N-1}^s = U_{2N}B_{2N}^s$ holds. By (\ref{Scanform1}) it follows that $U_{2N} = \idm$. One can now repeat this reasoning and see that $U_
 n, V_n =
  \idm$ for all $n > l$.\\
\\So the MPS (\ref{MPSSw1}) is gauge invariant iff for every $l = 1,\ldots, 2N$ there exist unitary matrices $U_l$ and $V_l$ (depending on $\varphi_l$) such that
\begin{equation}\label{condG} U_l^\dagger B_l^{s}V_l =e^{-i\varphi_l[\frac{s + (-1)^l}{2}]}B_l^s, C_{l-1}^{p}U_l = e^{-i\varphi_{l}p}C_{l-1}^{p} (l > 1), V_l^\dagger C_{l}^{p} = e^{i\varphi_{l}p}C_l^{p}.\end{equation}
Consider now the case $\varphi_l = 1$, then the matrices $U_l$ and $V_l$ do not depend on $\varphi_l$ anymore. The unitary matrices can be diagonalized (as exponential of a Hermitian matrix): $U_l = W_l^\dagger \Delta_{U_l} W_l$, $V_l = X_l^\dagger \Delta_{V_l}X_l,$ where $W_l, X_l$ are unitary matrices and $\Delta_{U_l}$ and $\Delta_{V_l}$ are diagonal matrices where all the diagonal-elements have modulus one. If we perform the following MPS-gauge transformation:
\begin{equation}B_l^s \rightarrow \bar{B}_l^s \equiv W_l B_l^s X_l^\dagger,  C_l^p \rightarrow \bar{C}_l^p \equiv X_l C_l^p W_{l+1}^\dagger,W_1 = W_{2N + 1} = 1 \end{equation}
the MPS (\ref{MPSSw1}) is unaffected and the conditions (\ref{condG}) now read
\begin{equation}\label{condGNew} \Delta_{U_l}^\dagger \bar{B}_l^{s}\Delta_{V_l} =e^{-i(s + (-1)^l)/2}\bar{B}_l^s, \bar{C}_{l-1}^{p}\Delta_{U_l} = e^{-ip}\bar{C}_{l-1}^{p} (l > 1), \Delta_{V_l}^\dagger \bar{C}_{l}^{p} = e^{ip}\bar{C}_l^{p}.\end{equation}
The property (\ref{Scanform1}) will also hold for $\bar{B}$ and $\bar{C}$, however the property (\ref{Scanform2}) is modified in the sense that $l_n^B$ and $l_n^C$ are not diagonal anymore (but they remain positive definite). We will denote this matrices with $l_n^{\bar{B}}$ and $l_n^{\bar{C}}$. As already mentioned, the entries of the diagonal matrices $\Delta_{U_l}$ and $\Delta_{V_l}$ are complex phase factors. Let $e^{-i\lambda_{l,j}}$, $j = 1, \ldots, n_{U_l}$, respectively $e^{-i\mu_{l,j}}$, $j = 1, \ldots, n_{v_l}$  be the eigenvalues of $\Delta_{U_l}$ with multiplicity $m(\lambda_{l,j})$ respectively of $\Delta_{V_l}$ with multiplicity $m(\mu_{l,j})$,
\begin{equation}\label{EqUV}\Delta_{U_l} = \sum_{j = 1}^{n_{u_l}}\sum_{\alpha_j = 1}^{m(\lambda_{l,j})}e^{-i\lambda_{l,j}}\vert \lambda_{l,j}, \alpha_j \} \{ \lambda_{l,j} ,\alpha_j \vert ,\Delta_{V_l} = \sum_{j = 1}^{n_{v_l}}\sum_{\alpha_j = 1}^{m(\mu_{l,j})}e^{-i\mu_{l,j}}\vert \mu_{l,j}, \alpha_j \} \{ \mu_{l,j} ,\alpha_j \vert ,\end{equation}
then we can write $\bar{B}$ and $\bar{C}$ as
\begin{subequations}
\begin{align}\bar{B}_l^s = \sum_{j = 1}^{n_{u_l}}\sum_{k = 1}^{n_{v_l}}\sum_{\alpha_j = 1}^{m(\lambda_{l,j})}\sum_{\beta_k = 1}^{m(\mu_{l,k})}
[\bar{B}_l^s]_{(\lambda_{l,j},\alpha_j);(\mu_{l,k},\beta_k)}\vert \lambda_{l,j} ,\alpha_j \} \{ \mu_{l,k} ,\beta_k \vert, l > 1  \label{EqBketform}\end{align}
\begin{align}\bar{C}_l^p = \sum_{j = 1}^{n_{v_l}}\sum_{k = 1}^{n_{u_{l+1}}}\sum_{\alpha_j = 1}^{m(\mu_{l,j})}\sum_{\beta_k = 1}^{m(\lambda_{l+1,k})}
[\bar{C}_l^p]_{(\mu_{l,j},\alpha_j);(\lambda_{l+1,k},\beta_k)}\vert \mu_{l,j} ,\alpha_j \} \{ \lambda_{l+1,k} ,\beta_k \vert, l < 2N  \label{EqCketform}\end{align}
\begin{align}\bar{B}_1^s = \sum_{k = 1}^{n_{v_1}}\sum_{\beta_k = 1}^{m(\mu_{1,k})}
[\bar{B}_1^s]_{1;(\mu_{1,k},\beta_k)}\{ \mu_{1,k} ,\beta_k \vert,\bar{C}_{2N}^p = \sum_{j = 1}^{n_{v_{2N}}}\sum_{\alpha_j = 1}^{m(\mu_{2N,j})}
[\bar{C}_{2N}^p]_{(\mu_{2N,j},\alpha_j);1}\vert \mu_{2N,j} ,\alpha_j \} .  \end{align}
\end{subequations}
Using (\ref{condGNew}) it follows that
\begin{equation}(e^{-i(p - \lambda_{l+1,k})} - 1)[\bar{C}_l^p]_{(\mu_{l,j},\alpha_j);(\lambda_{l+1,k},\beta_k)} = 0,(e^{-i(p - \mu_{l,j})} - 1)[\bar{C}_l^p]_{(\mu_{l,j},\alpha_j);(\lambda_{l+1,k},\beta_k)} = 0. \end{equation}
\begin{equation}(e^{-i(p - \mu_{2N,j})} - 1)[\bar{C}_{2N}^p]_{(\mu_{2N,j},\alpha_j);1} = 0, \end{equation}
so
\begin{equation}\label{restrC}[\bar{C}_l^p]_{(\mu_{l,j},\alpha_j);(\lambda_{l+1,k},\beta_k)} = \delta_{p,\mu_{l,j}} \delta_{p,\lambda_{l+1,k} }[c_l^p]_{\alpha_j, \beta_k},[\bar{C}_{2N}^p]_{(\mu_{{2N},j},\alpha_j);1} = \delta_{p,\mu_{2N,j}} [c_{2N}^p]_{\alpha_j, 1},\end{equation}
Note that $\lambda_{l,j}$ and $\mu_{l,j}$ are only unique up to a multiple of $2\pi$. By writing  $\delta_{p,\mu_{l,j}}$ we mean that we must take for $\mu_{l,j}$ up to a multiple of $2\pi$ the value $p$. Of course this will not influence the eigenvalue $e^{-i\lambda_{l,j}}$.

Assume now that there would exist a $\lambda_{l+1, k_0}$ ($ l < 2N $) with $\lambda_{l+1, k_0} \neq p$, $\forall p \in \mathbb{Z}$. Then it follows by (\ref{restrC}) that
\begin{equation}[\bar{C}_l^p]_{(\mu_{l,j},\alpha_j);(\lambda_{l+1,k_0},\beta_{k_0})} = 0, \end{equation}
$\forall p \in \mathbb{Z}, \forall j = 1, \ldots, n_{v_l}, \forall \alpha_j = 1, \ldots, m(\mu_{l,j}).$ If we now consider the non-singular matrix $l^{\bar{C}}_l$, see (\ref{Scanform2}), then
\begin{equation}\Biggl(\sum_{p \in \mathbb{Z}}({\bar{C}^{p}_l})^\dagger l^{\bar{C}}_{l} \bar{C}^{p}_l\Biggl)_{(\lambda_{l + 1,k_0}, \alpha_{k_0}),(\lambda_{l + 1,k}, \beta_k)} = 0, \end{equation}
$\forall \alpha_{k_0} = 1, \ldots, m(\lambda_{l,k_0})$,$\forall k = 1, \ldots, n_{u_{l + 1}}$, $\forall \beta_k = 1, \ldots, m(\lambda_{\lambda_{l+1,k}})$. By (\ref{Scanform2}) this would mean that $l^{\bar{B}}_{l+1}$ has a zero-row and would be singular which is a contradiction because $l^{\bar{B}}_{l+1}$ is positive definite. As a consequence all the $\lambda_{l,k}$ are integers. In the same way, but now by using the condition (\ref{Scanform1}) one proves that all the $\mu_{l,j}$ are integers.

We can write (\ref{EqUV}) as
\begin{equation}\label{EqUV2}\Delta_{u_l} = \sum_{q \in \mathbb{Z}}\sum_{\alpha_q = 1}^{D_q^l}e^{-iq}\vert q, \alpha_q \} \{ q ,\alpha_q \vert, \Delta_{v_l} = \sum_{q \in \mathbb{Z}}\sum_{\alpha_q = 1}^{{D_q^{'l}}} e^{-iq}\vert q, \alpha_q \} \{q, \alpha_q \vert ,\end{equation}
and expand $\bar{B}$, $\bar{C}$:
\begin{equation}\bar{B}_l^s = \sum_{q,r \in \mathbb{Z}}\sum_{\alpha_q = 1}^{D_q^l}\sum_{\beta_r = 1}^{D_r^{'l}}[\bar{B}_l^s]_{(q,\alpha_q);(r, \beta_r)}\vert q ,\alpha_q \} \{ r ,\beta_r \vert , \bar{C}_l^p = \sum_{q,r \in \mathbb{Z}}\sum_{\alpha_q = 1}^{D_q^{'l}}\sum_{\beta_r = 1}^{D_{r}^{l+1}}[\bar{C}_l^p]_{(q,\alpha_q);(r, \beta_r)}\vert q ,\alpha_q \} \{ r ,\beta_r \vert. \end{equation}
\begin{equation}\bar{B}_1^s = \sum_{r \in \mathbb{Z}}\sum_{\beta_r = 1}^{D_r^{'l}}[\bar{B}_1^s]_{1;(r, \beta_r)}\{ r ,\beta_r \vert , \bar{C}_{2N}^p = \sum_{q \in \mathbb{Z}}\sum_{\alpha_q = 1}^{D_q^{'2N}}[\bar{C}_l^p]_{(q,\alpha_q);1}\vert q ,\alpha_q \} \end{equation}
where $D_q^l$ respectively $D_q^{'l}$ denotes the multiplicity of the eigenvalue $q$ in the matrix $u_l$ respectively $v_l$. Note that $D^l = \sum_q D_q^l$ and $D^{'l} = \sum_q D_q^{'l}$. We have already proven, see (\ref{restrC}), that
\begin{equation}\label{eqCGaugeform}[\bar{C}_l^p]_{(q,\alpha_q);(r, \beta_r)} = \delta_{q,p}\delta_{q,r}[c_l^p]_{\alpha_q,\beta_r} , [\bar{C}_l^p]_{(q,\alpha_q);1} = \delta_{q,p}[c_l^p]_{\alpha_q,1} ,\end{equation}
where $c_l^p \in \mathbb{C}^{D_p^{'l} \times D_p^{l+1}}$. Finally, if we substitute (\ref{EqUV2}) in (\ref{EqBketform}), we obtain
\begin{equation} (e^{-i[(s+ (-1)^l)/2 + q - r]} - 1)[\bar{B}_l^s]_{(q,\alpha_q);(r, \beta_r)} = 0, (l>1), (e^{-i[(s - 1)/2 - r]} - 1)[\bar{B}_l^s]_{1;(r, \beta_r)} = 0\end{equation}
meaning that
\begin{equation}\label{eqBGaugeform}[\bar{B}_l^s]_{(q,\alpha_q);(r, \beta_r)} = \delta_{r, q + (s+(-1)^l)/2}[b_{l,q}^s]_{\alpha_q, \beta_r}, [\bar{B}_1^s]_{1;(r, \beta_r)} = \delta_{r, (s-1)/2}[b_{1,0}^s]_{1, \beta_r} \end{equation}
where $b_{l,q}^s \in \mathbb{C}^{D_q^l \times D_{q + (s + (-1)^l)/2}^{'l}}$ is random.

We have now proven that every MPS  that is invariant under local gauge transformations with $\varphi_l = 1$ can be brought in the form (\ref{eqCGaugeform}) and (\ref{eqBGaugeform}) by a MPS gauge transformation. A state in this form is also invariant under any gauge transformation.  Indeed, according to (\ref{condGNew}), we need to find unitary matrices $U_l$ and $V_l$ such that
\begin{equation}e^{-i\varphi_l p}\bar{C}_{l-1}^{p} =  \bar{C}_{l-1}^{p}U_l,e^{i\varphi_l p}\bar{C}_l^{p} = V_l^\dagger \bar{C}_l^{p},  e^{-i\varphi_l (s + (-1)^l)/2}\bar{B}_{l}^{s} = U_l^\dagger \bar{B}_{l}^{s}V_l,\end{equation}
where $\bar{B}$ equals  (\ref{eqBGaugeform}) and $\bar{C}$ equals (\ref{eqCGaugeform}). Taking
\begin{equation}[U_l]_{(q,\alpha_q);(r,\beta_r)} = \delta_{q,r}\delta_{\alpha_q,\beta_r}e^{-i\varphi_l q}, [V_l]_{(q,\alpha_q);(r,\beta_r)} = \delta_{q,r}\delta_{\alpha_q,\beta_r}e^{-i\varphi_l q},\end{equation}
solves this problem. This proves that every gauge invariant state can be brought in the form (\ref{eqCGaugeform}) and (\ref{eqBGaugeform}) by a MPS-gauge transformation and, conversely, that every MPS in the form (\ref{eqCGaugeform}) and (\ref{eqBGaugeform}) is gauge invariant.

\section{S2: Details on the implementation of the equilibrium simulations}
\subsection{A MPS ansatz for $CT$ invariant systems in the thermodynamic limit}
\noindent Consider a one-dimensional lattice of size $4N$ where every site $n$, $n \in \{-2N + 1, \ldots, 2N\}$, contains a $d-$dimensional Hilbert space $\hstates_n$ spanned by the basis $\{ \kets{q_n} : q_n = 1, \ldots, d \}$. The total Hilbert space is spanned by $\{ \ket{\bm{q}} \equiv \ket{q_{-2N+1} \ldots q_{2N} } : q_n = 1, \ldots, d\} \}$. We will take the thermodynamic limit ($N \rightarrow + \infty$). Let $H$ be a Hamiltonian which can be written as
\begin{equation}\label{HC}H = \sum_{n\in \mathbb{Z}}(\bm{C}T)^n\sum_{k = 1}^m\Bigl(h_1^{(k)}\otimes (C h_2^{(k)}C)\Bigl)(\bm{C}T)^{-n}, \end{equation}
where $h_i^{(k)}$ has only support on one site ($i = 1,2; k = 1,\ldots, m$), $\bm{C} = \otimes_{n \in \mathbb{Z}}C$, $C$ is an idempotent Hermitian operator which induces a permutation $c$ on the basis vectors ($C\kets{q_n} = \kets{c(q_n)}, c^2 = \idm$)  and $T$ is the translation operator. One can think of $C$ being the charge conjugation. It is clear that the Hamiltonian is invariant under the transformation $CT \equiv \bm{C}T:$ $H = (CT)H(CT)^\dagger$. Further one notes that the Hamiltonian is invariant under translations over an even number of sites. As a consequence it is possible to label the eigenstates of the Hamiltonian by the quantum numbers $k \in [ - \pi, \pi )$ and $\gamma \in \{-1, +1 \}$: $H\ket{k,\gamma} = E_{k,\gamma}\ket{k,\gamma}$ where $CT\ket{k,\gamma} = \gamma e^{-ik/2}\ket{k,\gamma}, T^2 \ket{k,\gamma} = e^{-ik}\ket{k,\gamma}$. The number $k$ corresponds to the momentum of the states (for translations over two sites). States with quantum number $\gamma
 = +1$ wi
 ll be referred to as \textit{scalar} particles and the excitations with quantum number $\gamma = -1$ will be referred to as \textit{vector} particles.\\
\\When the ground state of the Hamiltonian $H$ does not suffer from spontaneous symmetry breaking of the $CT$-symmetry, we can write down an ansatz which resembles a uniform MPS \cite{Nachtergaele} but is $CT$ invariant instead of translation invariant. Thereto we define $\ket{\bm{q}^c} \equiv \ket{c(q_{-2N+1}), q_{-2N + 2}, \ldots, c(q_{2n-1}), q_{2n}, \ldots, c(q_{2N-1}), q_{2N} }$ ($N \rightarrow + \infty$) which can be obtained by letting $C$ act on the odd components of $\ket{\bm{q}}$. The MPS $\ket{\Psi(A)}$ is now defined as
\begin{equation}\label{cuans}\ket{\Psi(A)} \equiv \sum_{\{q_n\}_{n \in \mathbb{Z}}}\bm{v}_L^\dagger \Biggl\{\prod_{n \in \mathbb{Z}}A^{q_n}\Biggl\} \bm{v}_R \ket{\bm{q}^c},  A \in \mathbb{C}^{D \otimes d \otimes D},\bm{v}_R, \bm{v}_L \in  \mathbb{C}^{D \times 1}.\end{equation}
One easily verifies that this state is $CT$ invariant and that it can be obtained from the uniform MPS
\begin{equation}\ket{\Psi^u(A)} \equiv \sum_{\{q_n\}_{n \in \mathbb{Z}}} \bm{v}_L^\dagger \Biggl\{\prod_{n \in \mathbb{Z}}A^{q_n}\Biggl\} \bm{v}_R \ket{\bm{q}},\end{equation}
by letting $C$ act on the odd sites.
\\ To obtain the ground state of $H$ we note that
\be \frac{\bra{\Psi(A^*)} H \ket{\Psi(A)}}{\bra{\Psi(A^*)}\Psi(A) } =\frac{\bra{\Psi^u(A^*)} H^u \ket{\Psi^u(A)}}{\bra{\Psi^u(A^*)}\Psi^u(A)\rangle },  \ee
where
\be H^u = \sum_{n\in \mathbb{Z}}T^n\sum_{k = 1}^m\Bigl(h_1^{(k)}\otimes h_2^{(k)} \Bigl)T^{-n}. \ee
This means that we have to find the ground state of the translational invariant Hamiltonian $H^u$ where we take as ansatz $\ket{\Psi^u(A)}$. The $A$ that we will obtain as the tensor corresponding to the ground state $\ket{\Psi^u(A)}$ of $H^u$ will also correspond to the ground state $\ket{\Psi(A)}$ of $H$. \\
\\Once we have obtained (to sufficient accuracy) the ground state $\ket{\Psi(A)}$ corresponding to the ground state energy density $E_0$ one can look for the excited states. For one-particle excited states an ansatz with quantum numbers $k$ and $\gamma$ ($k \in [ - \pi, \pi )$, $\gamma \in \{-1, +1 \}$) is \cite{HaegemanEA, HaegemanBMPS}
\begin{align}\label{eq TPVk cuMPS}\ket{\Phi_{k,\gamma}(B, A)}  =  \sum_{n \in \mathbb{Z}}\gamma^ne^{i(k/2)n} \sum_{\{q\} = 1}^d\bm{v}_L^\dagger \Biggl[\prod_{m < n }A^{q_m}\Biggl] B^{q_n}\Biggl[\prod_{m > n }A^{q_m}\Biggl]\bm{v}_R\ket{\bm{q^c}}.\end{align}
It is not hard to see that $CT\ket{\Phi_{k,\gamma}(B, A)}  = \gamma e^{-ik/2}\ket{\Phi_{k,\gamma}(B, A)}$ and $T^2\ket{\Phi_{k,\gamma}(B, A)} = e^{-ik}\ket{\Phi_{k,\gamma}(B, A)}$. These states can be obtained by letting $C$ act on the odd sites of the states $\ket{\Phi_{([k+(1-\gamma)\pi]/2)}^u(B, A)} $ where
\begin{align}\ket{\Phi_{l}^u(B, A)}  =  \sum_{n \in \mathbb{Z}}e^{iln} \sum_{\{q\} = 1}^d\bm{v}_L^\dagger \Biggl[\prod_{m < n }A^{q_m}\Biggl] B^{q_n}\Biggl[\prod_{m > n }A^{q_m}\Biggl]\bm{v}_R \ket{\bm{q}}, \forall l \in [-\pi, \pi[.\end{align}
The states $\ket{\Phi_{l}^u(B, A)}$ were introduced in \cite{HaegemanEA} as ansatz for momentum-$l$ particles for translational invariant systems.
\\To find the excited states we will apply the Rayleigh-Ritz method and find $B$ in such a way that it minimizes $\langle \Phi_{k,\gamma} (B^*,A^*)\vert H\ket{\Phi_{k,\gamma}(B, A)}/\langle \Phi_{k,\gamma}  (B^*,A^*)\ket{\Phi_{k,\gamma}(B, A)}$. By noting that
\begin{equation}\frac{\langle \Phi_{k,\gamma} (B^*,A^*)\vert H \ket{\Phi_{k,\gamma}(B, A)}}{\langle \Phi_{k,\gamma} (B^*,A^*)\ket{\Phi_{k,\gamma}(B, A)}}= \frac{\langle \Phi_{([k+(1-\gamma)\pi]/2)}^u(B^*,A^*)\vert H^u\ket{\Phi_{([k+(1-\gamma)\pi]/2)}^u(B, A)}}{\langle \Phi_{([k+(1-\gamma)\pi]/2)}^u  (B^*,A^*)\ket{\Phi_{([k+(1-\gamma)\pi]/2)}^u(B, A)}} \end{equation}
this problem is mapped to an analogue problem for uniform MPS. In \cite{HaegemanEA, HaegemanBMPS} it is discussed how to apply the Rayleigh-Ritz method to approximate excitations within the class of such states. \\
\\For the Schwinger model, the Hamiltonian reads
\begin{equation}H =  \frac{g}{2\sqrt{x}}\sum_{n \in \mathbb{Z}}(CT)^n\Biggl(L(1)^2 + \frac{\mu}{2} \bigl(\sigma_z(1) +1 ) + x (\sigma^-(1) e^{-i\theta(1)}[C\sigma^-(2)C] + h.c.\bigl)\Biggl)(CT)^{-n}, \end{equation}
where $C$ is the charge conjugation: $C\ket{s,q} = \ket{-s, -q}$, implying that
\begin{equation}H^u =  \frac{g}{2\sqrt{x}}\sum_{n \in \mathbb{Z}}{T}^n\Biggl(L(1)^2 + \frac{\mu}{2} \bigl(\sigma_z(1) +1 ) + x (\sigma^-(1) e^{-i\theta(1)}\sigma^-(2) + h.c.\bigl)\Biggl)T^{-n}. \end{equation}

\subsection{Special features of gauge invariant MPS}
We will now construct an ansatz of the form (\ref{cuans}) which is gauge invariant. We start from a MPS invariant under translations over an even number of sites and perform a charge conjugation on the odd sites:
\begin{equation}\label{ansatzCunif} \sum_{\{s_n\} = \pm 1}\sum_{\{p_n\} \in  \mathbb{Z}} \bm{v}_L^\dagger \Biggl\{\prod_{n = -N+1}^N B_1^{-s_{2n-1}}C_1^{-p_{2n-1}}B_2^{s_{2n}}C_2^{p_{2n}}\Biggl\} \bm{v}_R \ket{\{s_{2n-1},p_{2n-1},s_{2n},p_{2n} \}},\end{equation}
where $N \rightarrow + \infty$. To make the state gauge invariant, we will require that they have the form (\ref{eqCGaugeform}) and (\ref{eqBGaugeform}):
\begin{equation} [B_l^{(-1)^ls}]_{(q,\alpha_q);(r, \beta_r)} = \delta_{r, q + (s+(-1)^l)/2}[b_{l,q}^s]_{\alpha_q, \beta_r}, [C_l^{(-1)^lp}]_{(q,\alpha_q);(r, \beta_r)} = \delta_{q,p}\delta_{q,r}[c_l^p]_{\alpha_q,\beta_r},\end{equation}
where $l = 1,2$. We will now perform the following MPS-gauge transformation: $B_1^s \rightarrow \bar{B}_1^s = UB_1^s$, $C_1^p \rightarrow \bar{C}_1^p = C_1^p$, $B_2^s \rightarrow \bar{B}_2^s = B_2^s$ and $C_2^p \rightarrow \bar{C}_2^p = C_2^pU^\dagger$ where $[U]_{(p,\alpha);(q,\beta)} = \delta_{p,-q}\delta_{\alpha,\beta}$. It follows that
\begin{equation}[\bar{B}_1^{s}\bar{C}_1^{p}]_{(q,\alpha);(r,\beta)} = b_{1,-q}^{-s}c_1^{-q - (s +1)/2}\delta_{p, q + (s+1)/2}\delta_{r, -q - (s+1)/2}, \end{equation}
\begin{equation}[\bar{B}_2^{s}\bar{C}_2^{p}]_{(q,\alpha);(r,\beta)} = b_{2,q}^{s}c_2^{q + (s +1)/2}\delta_{p, q + (s+1)/2}\delta_{r, -q - (s+1)/2}. \end{equation}
By taking $b_{1,-q}^{-s} = b_{2,q}^s \equiv b_{q}^{s}$, $c_{1}^{-p} = c_2^{p} \equiv c^p$ and defining
\be\label{GIA2} [A^{s,p}]_{(q,\alpha_q);(r,\beta_r)} = [a^{s,p}]_{\alpha_q,\beta_r}\delta_{p, q + (s+1)/2}\delta_{r, -q - (s+1)/2} \ee
where $a^{s,p} = b_{p - (s+1)/2}^s \,c^{p}$, the state (\ref{ansatzCunif}) can be written as
\begin{equation}\ket{\Psi(A)} =  \sum_{\{s_n\} = \pm 1}\sum_{\{p_n\} \in  \mathbb{Z}}  \bm{v}_L^\dagger \Biggl\{\prod_{n = -2N+1}^{2N}A^{s_n,p_n}\Biggl\} \bm{v}_R \ket{\{-s_{2n-1},-p_{2n-1},s_{2n},p_{2n} \}},\end{equation}
which is a gauge and $CT$ invariant ansatz.\\
\\As already mentioned, the ground state can be obtained by minimizing
\be \frac{\bra{\Psi^u(A^*)} H^u \ket{\Psi^u(A)}}{\bra{\Psi^u(A^*)}\Psi^u(A)\rangle }.\ee
We did this by imaginary time evolution using the TDVP method \cite{HaegemanBMPS, HaegemanTDVP}. The TDVP method evolves the Schr\"odinger equation (SE), $i \partial_t \vert \Psi^u\bigl(A\bigl) \rangle = H^u\vert \Psi^u\bigl(A\bigl) \rangle$, within the manifold of uMPS. To this end the right-hand side of the SE is replaced by $\vert \Phi^u \bigl(B_{H}(A),A\bigl)\rangle$, where $\ket{\Phi^u(B,A)}$ is given by
\bea\sum_{m\in\mathbb{Z}}\sum_{q_n}\bm{v}_L^\dagger \left(\prod_{n < m}A^{q_n}\right)B^{q_m}\left(\prod_{n > m}A^{q_n}\right) \bm{v}_R\ket{\bm{q}}, \eea
$B^{q}$ has also the block structure (\ref{GIA2}), $B_{{H}}(A) = \mbox{arg}\min_B \bigl\vert \bigl\vert \vert \Phi^u(B,A) \rangle - H\vert\Psi^u(A)\rangle \bigl\vert\bigl\vert$ and $\vert \Phi^u(B_{H}(A) ,A) \rangle \perp \vert\Psi^u(A) \rangle$. This last condition is imposed to have norm-preservation up to first order in the time step. The SE will boil down to an ordinary differential equation for the variational parameters $a^{s,p}$ of the form $i\dot{a} = b_{H}(a)$, where $b_{H}(a)$ can be computed in $\mathcal{O}\bigl((2p_{max} + 1) \max_p D_p^3\bigl)$ time. For the explicit formulas in the non gauge invariant case we refer to \cite{HaegemanTDVP,HaegemanBMPS}, the expressions in our case are obtained by taking into account the block-structure (\ref{GIA2}). Starting from a random state $\ket{\Psi(A)}$ we will then evolve towards the ground state with imaginary time $\tau = it$ and stop once $\sqrt{\bra{\Phi^u \bigl(B_{H}^*(A),A^{*}\bigl)}\Phi^u \bigl(B_{H}(A),A\bigl) \rangle / \vert\mathbb{Z}\vert }$ is below a certain tolerance. In our simulations we took this tolerance equal to $10^{-9}$.\\
Once we have obtained the ground state, we can use the ans\"atze (\ref{eq TPVk cuMPS}) to approximate the excited states. If we put $[B^{s,p}]_{(q,\alpha_q);(r,\beta_r)} = [b^{s,p}]_{\alpha_q,\beta_r}\delta_{p, q + (s+1)/2}\delta_{r, -q - (s+1)/2}$ they will automatically be gauge invariant. The variational freedom then lies within the matrices $b^{s,p}$. We refer to \cite{HaegemanEA,HaegemanBMPS} for the explicit formulas. Note that in (\ref{eq TPVk cuMPS}) the energy corresponding to the physical momentum $k$ is obtained by replacing $k \rightarrow 2ka$. The ansatz thus reads
\begin{align}\label{eq TPVk cuMPS2}\ket{\Phi_{k,\gamma}(B, A)}  =  \sum_{n \in \mathbb{Z}}\gamma^ne^{ikna} \sum_{\{q\} = 1}^d\bm{v}_L^\dagger \Biggl[\prod_{m < n }A^{q_m}\Biggl] B^{q_n}\Biggl[\prod_{m > n }A^{q_m}\Biggl]\bm{v}_R\ket{\bm{q^c}}.\end{align}

\section{S3: Numerical results for groundstate and one-particle excitations}

The continuum limit $a\rightarrow 0$ of the Schwinger model corresponds to the limit $x\rightarrow\infty$ in Eq. (2) (in Main Text). To obtain the energies of the ground state and of the one-particle excitations in this limit, we have calculated these quantities for values of $x=100,200,300,400,600,800$. At every $x$ we considered different values of $D$ till convergence was reached at some $D_{max}$. We estimated the truncation error on $D$ from comparison of the result for $D=D_{max}$ with the result for the next to largest value of $D$. Larger values of $x$ typically required larger values of $D$ for the same order of the error. For instance for $m/g=0.5$ our maximal $D$ varied from 185 for $x=100$ to 358 for $x=800$. This scaling of $D$ is not surprising, as it is well known that MPS representations require larger $D$ for systems with larger correlation lengths $\xi$ (in units of the lattice spacing) \cite{Cirac}. For the Schwinger model $\xi$ indeed diverges in the $x\rightarrow \infty$ limit.

To extrapolate towards $x\rightarrow \infty$ we used a third order polynomial fit in $1/\sqrt{x}$ through the largest five $x$-values. Similar to \cite{Byrnes} our extrapolation error is then estimated by considering a third and fourth order polynomial through all six points, taking the error to be the maximal difference with the original inferred value.

In table 1 we display our resulting values for the ground state energy density and the mass of the different one-particle excitations. For $m/g=0$ this can be compared with the exact result that follows from bosonization \cite{Schwinger1}. In this limit the model reduces to a free theory, of one bosonic vector ($\gamma=-1$) particle with mass $M_{v,1} = 1/\sqrt{\pi}=0.56419$ and with a ground state energy density $\omega_0=-1/\pi =-0.318310$ (both in units $g=1$).

Furthermore, in the strong coupling expansion $g/m\gg 1$ on this exact result, it is found that the vector boson becomes an interacting particle, leading to two more stable bound states.  There appears one scalar boson that is a stable bound state of two vectors and one more vector boson, that is best interpreted as a bound state of the scalar and the original lowest mass vector \cite{Coleman,Adam}. For $g/m\neq 0$ we also find three excited states, one scalar and two vectors, with the hierarchy of masses $M_{v,1}<M_{s,1}<M_{v,2}$ matching that of the strong coupling result. But notice that for our values of $g/m$, the strong coupling expansion result is not reliable anymore, making a quantitative comparison useless. One can also show that in the continuum limit the ground state energy is independent of $g/m$ which is compatible with our findings.

As explained in the text we truncate the charges $p$ (the eigenvalues of $L$) at $p_{max} = 3$ which corresponds to taking the bond dimensions $D_q = 0$ for $\vert q \vert > p_{max}$. Physically, this truncation hinges on the fact that the first term in the Hamiltonian (2) (in Main Text) ($\varpropto \sum_n L^2(n)$) punishes states with large charges. As a consequence we expect such states not to be relevant for the low-energy physics at strong coupling. In fig.1a in the main text we illustrate how one can check this assumption and determine the proper truncation by looking at the relative weight of the different charge sectors. As an extra check on our truncation we have performed another simulation, now with $p_{max} = 4$, again for $x = 100, m/g = 0.25$.  In fig.3a we plot again the Schmidt coefficients for the ground state and by comparing this figure with the figure in the main text, we clearly see that we can indeed neglect the contributions from the $p = \pm 4$ charge sectors. Furthermore, we can compare the approximations of the energies of our excited states. For instance for $m/g
 = 0.25$ and $x = 100$ we obtain $M_{v,1} = 1.04206770$ $M_{s,1} = 1.7515838$ $M_{v,2} = 2.3570578$ for $p_{max} = 4$ and $M_{v,1} = 1.04206777$ $M_{s,1} = 1.7515839$ $M_{v,2} = 2.3570577$ for $p_{max} = 3$. The absolute difference is only of order $10^{-7}$ indicating again that the charge sectors $p =  \pm 4$ can indeed be ignored.

Finally, as mentioned in the text, a nice cross-check of our method follows from calculating the excitation energies for non-zero momenta $k$. The Schwinger model is Lorentz invariant in the continuum limit, so we should have an approximate Einstein dispersion relation at finite lattice spacing $a$, for small momenta $ka\ll 1$. As shown in fig.3b, this is precisely what we find.

\begin{figure}
\begin{tabular}{rr}
  \includegraphics[width=80mm]{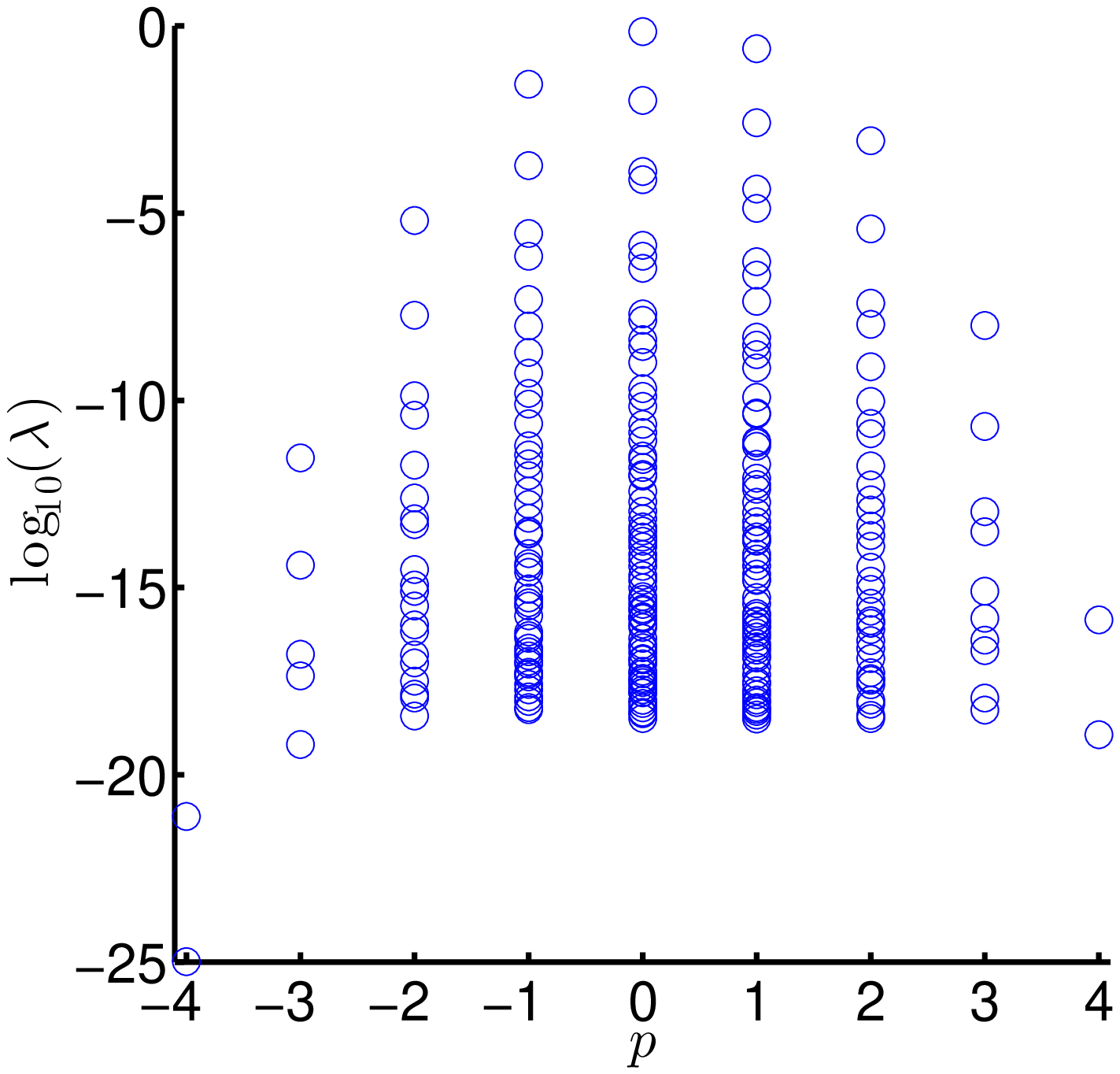} &   \includegraphics[width=84mm]{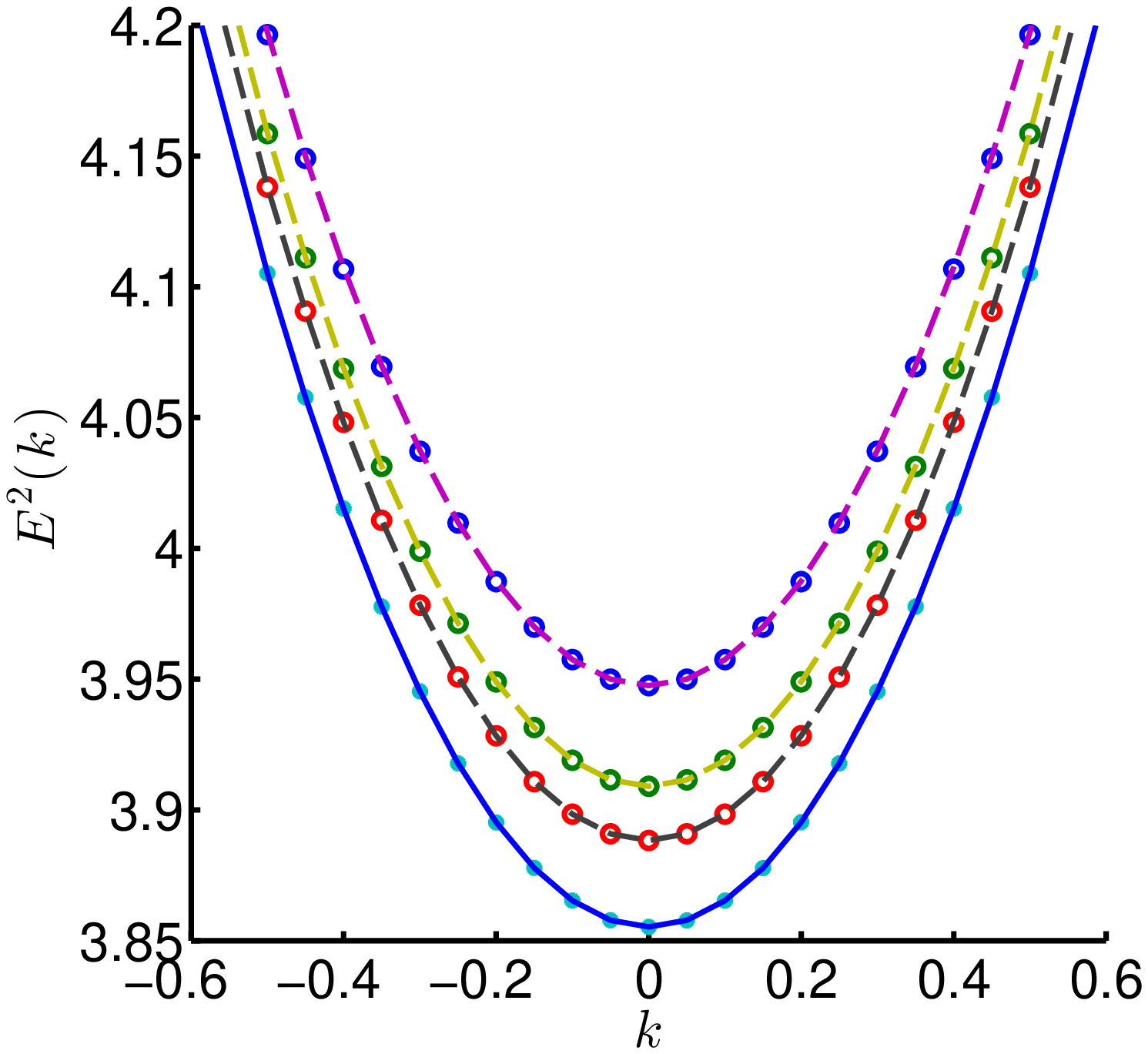}
\end{tabular}
\caption{\label{fig:ResA1} Left (a): distribution of the (base-10) logarithm of the Schmidt coefficients in every charge sector $m/g = 0.25$, $x = 100$ and $p_{max} = 4$. Right (b): Fit of the Einstein-dispersion relation $E^2 = k^2 + M_{v,1}^2(x)$ for $m/g = 0.75$, $x = 100, 300, 800$ (dashed lines) to the data (small circles). The stars represent the estimated continuum values, the full line (lowest lying curve) is the curve $E^2 = k^2 + M_{v,1}^2$.  }
\end{figure}

\section{S4: Real-time evolution with iTEBD}\label{app4}
We recall that the iTEBD algorithm \cite{iTEBD} expands the operator $\exp(-iHdt)$, $H = \sum_{n \in \mathbb{Z}}h_{n,n+1}$, through a Suzuki-trotter decomposition \cite{TrotterSuzuki} as a sequence of two-site gates $U_{n,n+1}(dt') = \exp(-ih_{n,n+1}dt')$ ($dt' \leq dt < 1$) which we rearrange into the gates $V_n = \bigotimes_{r \in \mathbb{Z}}U_{2r+n,2r+1+n}$, $(n = 1,2)$. In our case we used a fourth-order Trotter-expansion:
\be\exp(-iHdt) = V_1(s dt/2)V_2(sdt)V_1((1-s)/2 dt)V_2((1-2s)dt)V_1((1-s)/2 sdt) V_2(sdt)V_1(sdt/2) + \mathcal{O}(dt^5),\ee
where $s =  1/(2 - \sqrt[3]{2})$. Applying such a Trotter-gate $V_n$ to
\be\label{itebdform} \ket{\Psi(A_1,A_2)} = \sum_{q_n} \bm{v}_L^\dagger \left(\prod_{n \in \mathbb{Z}}A_1^{q_{2n-1}}A_2^{q_{2n}}\right) \bm{v}_R \ket{\{{(q_{2n-1},q_{2n})\}}}, A_n^{q_n} \in \mathbb{C}^{D_n \times D_{n+1}} \ee
results in
\be V_n\ket{\Psi(A_1,A_2)} = \sum_{q_n} \bm{v}_L^\dagger \left(\prod_{n \in \mathbb{Z}}B_{n,n+1}^{q_n,q_{n+1}}\right) \bm{v}_R \ket{\{{(q_{2n-1},q_{2n})\}}} \ee
where $q_n = (s_n,p_n)$, $s_n = \pm 1, p_n \in \mathbb{Z}[p_{min}^{n+1},p_{max}^{n+1}]$ and
\be \label{eqBn}B_{n,n+1}^{q_1,q_2} = \sum_{q'_1,q'_2}[U_{n,n+1}]_{(q_1,q_2);(q'_1,q'_2)}A_{n}^{q'_1}A_{n+1}^{q'_2}. \ee
Note that in our case $U_{n,n+1}$ and $B_{n,n+1}$ only depend on the parity of $n$. From now on our equations for $n$ have to be read modulo 2 for $n \in \{1,2 \}$.  In order to reobtain a MPS of the form $\ket{\Psi(\tilde{A}_1,\tilde{A}_2)}$ one performs a singular value decomposition (SVD) of $l^{1/2}B_{n,n+1}r^{1/2}$:
\be \label{svdiTEBD}l^{1/2}B_{n,n+1}^{q_n,q_{n+1}}r^{1/2} = U_n^{q_n}\Sigma_{n,n+1} W_{n+1}^{q_{n+1}},\ee
where $U_n^{q_n} \in \mathbb{C}^{D_n \times d_nD_{n}}, \Sigma_{n,n+1} \in \mathbb{C}^{d_{n} D_{n} \times d_{n+1}D_{n}}, W_{n+1}^{q_{n+1}}\in \mathbb{C}^{d_{n+1}D_{n} \times D_{n}}$ $(d_{n} = 2(p_{max}^{n+1} - p_{min}^{n+1} + 1))$ obey
\be \sum_{q}(U_n^q)^\dagger U_n^q = \idm, \sum_{q} W_{n+1}^q (W_{n+1}^q)^\dagger = \idm \ee
and $\Sigma_{n,n+1}$ is a diagonal matrix with non-negative elements in decreasing order.

Here, $l$ and $r$ are the left and right eigenvector of the transfer matrix corresponding to the eigenvalue with largest magnitude. Remember that the transfer matrix \cite{HaegemanTDVP,HaegemanBMPS} acts on the right on a matrix $r$ as $\sum_{q_n,q_{n+1}}A_n^{q_n}A_{n+1}^{q_{n+1}}r{A_{n+1}^{q_{n+1}}}^\dagger {A_n^{q_n}}^\dagger$ and on the left on a matrix $l$ as $\sum_{q_n,q_{n+1}}{A_{n+1}^{q_{n+1}}}^\dagger {A_n^{q_n}}^\dagger l A_n^{q_n}A_{n+1}^{q_{n+1}}$. For an injective MPS the transfer matrix has a unique left and right eigenvector (say $l$ and $r$) corresponding to the eigenvalue with largest magnitude which is real. Moreover $l$ and $r$ can be taken positive definite, diagonal, and such that tr$(lr)$ = 1. One can also scale the tensors $A$ such that this largest eigenvalue equals one which implies that our state is normalized. From now on, we will assume that this is the case. The non-zero diagonal elements of $\Sigma_{n,n+1}\Sigma_{n,n+1}^\dagger$ are then the Schmidt coefficients associated to the bipartition $(-\infty, n] \cup [n + 1,+\infty)$ of our lattice.

We now approximate the SVD $l^{1/2}B_{n,n+1}^{q_n,q_{n+1}}r^{1/2} = U_n^{q_n}\Sigma_{n,n+1} W_{n+1}^{q_{n+1}}$ by its truncated version $l^{1/2}B_{n,n+1}^{q_n,q_{n+1}}r^{1/2} \approx U_n^{q_n}\tilde{\Sigma}_{n,n+1} W_{n+1}^{q_{n+1}}$, where $\tilde{\Sigma}_{n,n+1} \in \mathbb{C}^{d_nD_n \times d_{n+1}D_{n+1}}$ is the diagonal matrix which contains the $\tilde{D}_{n+1}$ diagonal elements of $\Sigma_{n,n+1}$ larger than a certain tolerance $\epsilon$ (in decreasing order) and has all the other diagonal elements zero. We can then perform the following decomposition: $\tilde{\Sigma}_{n,n+1} = \tilde{\Sigma}_{n}^{1/2}\tilde{\Sigma}_{n+1}^{1/2}$ where $\tilde{\Sigma}_{n}^{1/2}\in \mathbb{C}^{d_{n}D_{n}\times \tilde{D}_{n+1}}$ and $\tilde{\Sigma}_{n+1}^{1/2}\in \mathbb{C}^{\tilde{D}_{n+1} \times d_{n+1}D_{n}}$  are diagonal matrices which contain the $\tilde{D}_{n+1}$ square roots of the non-zero diagonal elements of $\tilde{\Sigma}_{n,n+1}$ on their diagonal. The iTEBD now approximates $V _n\ket{\Psi(A_1,A_2)}$ by $\ket{\Psi(\tilde{A}_1,\tilde{A}_2)}$ where $\tilde{A}_n^{q_n} = l^{-1/2}U_n^{q_n}\tilde{\Sigma}_{n}^{1/2}$ and $\tilde{A}_{n+1}^{q_{n+1}} = \tilde{\Sigma}_{n+1}^{1/2}W_{n+1}^{q_{n+1}}r^{-1/2}$. By this approximation the bond-dimension at site $n+1$ is then $\tilde{D}_{n+1}$ instead of $\min (d_{n},d_{n+1}) D_{n}$ but will introduce an error in the expectation values of the observables of order of the sum of the discarded diagonal elements of $\Sigma_{n,n+1}$ (i.e. of order $\epsilon$).\\
\\ If we impose gauge invariance,
\be\label{GIA4} [A_n^{s,p}]_{(q,\alpha_q);(r,\beta_r)} =  [a_n^{s,p}]_{\alpha_q, \beta_r}\delta_{p, q + (s+(-1)^n)/2} \delta_{r,p}, \ee
where
\be [a_n^{s,p}] \in \mathbb{C}^{D_{p - (s + (-1)^n)/2}^n \times D_p^{n+1}},D_q^n = 0 \mbox{ for }q \notin \mathbb{Z}[p_{min}^n,p_{max}^n],D_q^{n+1} = 0 \mbox{ for }q \notin \mathbb{Z}[p_{min}^{n+1},p_{max}^{n+1}], \ee
as follows from (\ref{eqCGaugeform}) and (\ref{eqBGaugeform}), one can check that for the Hamiltonian $H$,
\be\label{equationH2} H= \frac{g}{2\sqrt{x}}\Biggl( \sum_{n \in \mathbb{Z}} [{L}(n) + \alpha]^2 + \frac{\mu}{2} \sum_{n \in \mathbb{Z}}(-1)^n(\sigma_z(n) + (-1)^{n}) + x \sum_{n \in \mathbb{Z}}(\sigma^+ (n)e^{i\theta(n)}\sigma^-(n + 1) + h.c.)\biggl),\ee
$B_{n,n+1}$, see (\ref{eqBn}), equals
\be [B_{n,n+1}^{s_1,p_1,s_2,p_2}]_{(q,\alpha_q);(r,\beta_r)} = b_{n,n+1}^{p_1,s_1,s_2}\delta_{p_1,q + (s_1 + (-1)^n)/2}\delta_{p_2,p_1+ (s_2 + (-1)^{n+1})/2}\delta_{p_2,r} \ee
where
\be b_{n,n+1}^{p,s_1,s_2} = \sum_{t_1,t_2 = \pm 1}\delta_{s_1 + t_1, s_2 + t_2}[U_{n,n+1}^{p}]_{(s_1,s_2);(t_1, t_2)}a_n^{t_1,p + (s_2 - t_2)/2}a_{n+1}^{t_2,p + (s_2 + (-1)^{n+1})/2} \ee
and
\be U_{n,n+1}^{p}(dt) = \exp\Biggl[\frac{-i}{2\sqrt{x}}\Bigl([p + \alpha]^2\idm\otimes \idm + (-1)^n \frac{\mu}{2} \sigma_z(n)\otimes \idm + x [\sigma^+(n)\otimes \sigma^{-}(n+1) + h.c.]\Bigl)dt\Biggl] \in \mathbb{C}^{(2\otimes 2)\times (2 \otimes 2)}.\ee
In this case, as can be checked, the left and right eigenvector, $l$ and $r$, corresponding to the largest eigenvalue of the transfer matrix in magnitude will also have a block-structure, $[l]_{(p,\alpha);(q,\beta)} = [l^{p}]_{\alpha,\beta} \delta_{p,q}, [r]_{(p,\alpha);(q,\beta)} = [r^{p}]_{\alpha,\beta} \delta_{p,q}$, where $l^p$ and $r^p$ can be taken positive definite and diagonal. For every $p$ for which $p - (s_1 + (-1)^n)/2, p + (s_2 + (-1)^{n+1})/2 \in \mathbb{Z}[p_{min}^{n},p_{max}^{n}]$ we perform a  decomposition similar to (\ref{svdiTEBD}):
$$\bigl[l^{p - (s_1 + (-1)^n)/2}\bigl]^{1/2}\; b_{n,n+1}^{p,s_1,s_2} \; \bigl[r^{p + (s_2 + (-1)^{n+1})/2}\bigl]^{1/2} = U_n^{p,s_1}\Sigma_{n,n+1}^{p}W_{n+1}^{p,s_2}$$
where $U_n^{p,s} \in \mathbb{C}^{D_{p - (s + (-1)^n)/2}^{n} \times d_{p}^{n+1}}, \Sigma_{n,n+1}^{p} \in \mathbb{C}^{d_{p}^{n+1} \times d_{p}^{n+1}}, W_{n+1}^{p,s}\in \mathbb{C}^{d_{p}^{n+1} \times D_{p + (s + (-1)^{n+1})/2}^{n+1}}$ ($d_{p}^{n+1} = D_{p} + D_{p + (-1)^{n+1}}$) obey
\be \sum_{s = \pm 1}(U_n^{p,s})^\dagger U_n^{p,s} = \idm_{d_{p}^{n+1}}, \sum_{s = \pm 1} W_{n+1}^{p,s} (W_{n+1}^{p,s})^\dagger = \idm_{d_{p}^{n+1}} \ee
and $\Sigma_{n,n+1}^{p}$ is a positive definite and diagonal square matrix. By discarding the diagonal elements of $\Sigma_n^{p}$ smaller than a tolerance $\epsilon$, we can truncate the bond-dimension on site $n+1$ in every sector of the charge-representation. It is even possible to discard the charge $p$-representation if all singular values of $\Sigma^{p}$ are smaller than $\epsilon$. We also see that if we have truncated our charges between $p_{min}^n$ and $p_{max}^n$ at site $n$ (i.e. $D_p^n = 0$ for $p \notin \mathbb{Z}[p_{min}^n,p_{max}^n]$), that at site $n+1$ we allow charges in the interval $p_{min}^{n} - 1$ and $p_{max}^n$ (i.e. $D_p^{n+1} = 0$ for $p \notin \mathbb{Z}[p_{min}^n-1,p_{max}^n]$) if $n$ is odd and charges in the interval $p_{min}^{n}$ and $p_{max}^n + 1$ (i.e. $D_p^{n+1} = 0$ for $p \notin \mathbb{Z}[p_{min}^n,p_{max}^n + 1]$) if $n$ is even. In this way it is possible to dynamically increase our charges which is useful for larger values of the background electric field $\alpha$. $a_n$ and $a_{n+1}$ are now updated by $\tilde{a}_n$ and $\tilde{a}_{n+1}$ via the prescription
\be \tilde{a}_n^{s,p} = \bigl[l^{p - (s_+ (-1)^n)/2}\bigl]^{-1/2}U_n^{p,s}(\tilde{\Sigma}^{p})^{1/2}, \tilde{a}_{n+1}^{s,p} = (\tilde{\Sigma}^{p})^{1/2}W_{n+1}^{p,s} \bigl[r^{p + (s + (-1)^{n+1})/2}\bigl]^{-1/2},\ee
where $\tilde{\Sigma}^p$ contains the singular values of $\Sigma$ larger than $\epsilon$.\\
\\In our  ground state simulations we exploited $CT$ invariance to write the ground state as a uniform MPS in a basis with charge conjugation on the odd sites (see  (\ref{GIA2})), i.e:
\be\ket{\Psi(A)} =  \sum_{\{s_n\} = \pm 1}\sum_{\{p_n\} \in  \mathbb{Z}}  \bm{v}_L^\dagger \Biggl\{\prod_{n = -\infty}^{\infty}A^{s_n,p_n}\Biggl\} \bm{v}_R \ket{\{-s_{2n-1},-p_{2n-1},s_{2n},p_{2n} \}},\ee
where
\be \label{GIA3}[A^{s,p}]_{(q,\alpha_q);(r,\beta_r)} = [a^{s,p}]_{\alpha_q,\beta_r}\delta_{p, q + (s+1)/2}\delta_{r, -q - (s+1)/2}, a^{s,p}\in \mathbb{C}^{D_{p - (s + 1)/2}\times D{-p}}. \ee
If $U$ performs a charge flip on the virtual level, $[U]_{(p,\alpha);(q,\beta)} = \delta_{p,-q}\delta_{\alpha,\beta}$, we set
\be A_1^{s,p} = UA^{-s,-p} \mbox{ and }A_2^{s,p} = A^{s,p}U\ee
and find that (because $U^2 = \idm$)
\be \ket{\Psi(A)}  =   \sum_{q_n} \bm{v}_L^\dagger \left(\prod_{n \in \mathbb{Z}}A_1^{q_{2n-1}}A_2^{q_{2n}}\right) \bm{v}_R\ket{\{s_{2n-1},p_{2n-1},s_{2n},p_{2n} \}}.\ee
This brings $\ket{\Psi(A)}$ in the form (\ref{itebdform}) that we we can then use as starting point for the time evolution with iTEBD. Moreover it follows from (\ref{GIA3}) that
\be [A_1^{s,p}]_{(q,\alpha_q);(r,\beta_r)} = [a^{-s,-p}]_{\alpha_q,\beta_r}\delta_{p,q + (s-1)/2} \delta_{r,p} \mbox{ and }  [A_2^{s,p}]_{(q,\alpha_q);(r,\beta_r)} = [a^{s,p}]_{\alpha_q,\beta_r}\delta_{p,q + (s+1)/2} \delta_{r,p}. \ee
Our tensors are now brought in the form (\ref{GIA4}) where $a_1^{s,p} =a^{-s,-p}$ and $a_2^{s,p} = a^{s,p}$. For the bond dimensions we have that $D_1^p = D_{-p}$ and $D_2^p = D_p$.

\section{S5: Linear response regime}
Consider a Hamiltonian $H = H_0 + \lambda V$, an observable $O$ and a state $\ket{\Psi(t)}$. We evolve this state according to the Schr\"{o}dinger equation, $i\partial_t\ket{\Psi(t)} = H\ket{\Psi(t)}$, starting at time $t = 0$. Truncating the corresponding Dyson series at second order then gives:
\be \bra{\Psi(t)}O \ket{\Psi(t)} \approx \bra{\Psi_0}O \ket{\Psi_0} -i\lambda  \int_0^t dt'\bra{\Psi_0}[O_I(t),V_I(t')]\ket{\Psi_0} - \lambda^2 \int_0^t dt' \int_0^{t'} dt'' \bra{\Psi_0}[[O_I(t),V_I(t')], V_I(t'')]]\ket{\Psi_0}, \ee
with $\ket{\Psi_0} = \ket{\Psi(0)}$. Here, the subscript $I$ denotes the interaction picture: $O_I(t) = e^{iH_0t}Oe^{-iH_0t}$. Focussing on the case of the Schwinger model with a background $\alpha$ (\ref{equationH2}) we observe that applying this background field is equivalent to a perturbation $(\alpha/\sqrt{x}) V = (\alpha/\sqrt{x})\sum_n L(n)$ to the Hamiltonian (2) (in Main Text). So, in our case $H_0$ corresponds to the Hamiltonian (2) (in Main Text), $\lambda = \alpha/\sqrt{x}$ and $V = \sum_{n \in \mathbb{Z}}L(n)$, implying that (up to corrections of order $\mathcal{O}(t^3\alpha^3)$):
\be \bra{\Psi(t)}O \ket{\Psi(t)}= \bra{\Psi_0}O \ket{\Psi_0} -i\frac{\alpha}{\sqrt{x}}  \int_0^t dt'\bra{\Psi_0}[O_I(t),V_I(t')]\ket{\Psi_0} - \frac{\alpha^2}{x} \int_0^t dt' \int_0^{t'} dt'' \bra{\Psi_0}[[O_I(t),V_I(t')], V_I(t'')]]\ket{\Psi_0}, \label{dyson}\ee
where $V_I(t) = e^{iH_0t}\sum_n L(n) e^{-iH_0t}$.

In fig.2b in the main text we display our results for the time evolution of the $CT=-1$ electric field operator, $O=\f{1}{2|\mathbb{Z}|}\sum_n L(n)$. In that case the linear response term in (\ref{dyson}) can be expanded as (with $H_0\ket{\Psi_0}=E_0\ket{\Psi_0}$): \bea
-i\frac{\alpha}{\sqrt{x}}  \int_0^t dt'\bra{\Psi_0}[O_I(t),V_I(t')]\ket{\Psi_0}&=&-i\frac{\alpha}{2\sqrt{x}|\mathbb{Z}|}  \int_0^t dt'\bra{\Psi_0}\sum_n L(n) \left(\exp^{i(H_0-E_0)(t'-t)}\right)\sum_n L(n) \ket{\Psi_0}+c.c.\nonumber\\
&=& \frac{\alpha}{\sqrt{x}|\mathbb{Z}|}\sum_i\frac{\cos\left(\left(E_i-E_0\right) t\right)-1}{(E_i-E_0)}|\bra{\Psi_0}\sum_n L(n)\ket{\Psi_i}|^2\,,\label{expandlin}
\eea
where the $i$-summation is restricted to all $CT=-1$ eigenstates of the unquenched Hamiltonian $H_0$, since $\sum_n L(n) \stackrel{CT}{\rightarrow} - \sum_n L(n)$ and $CT\ket{\Psi_0}=\ket{\Psi_0}$.  (Notice also that for the continuous part of the spectrum this formal sum should be read as an integral with the proper measure.)  From the linear $\alpha$-scaling we can infer that the $\alpha=0.005$ and $\alpha=0.01$ result in fig.2b (Main Text) remain in the linear response regime throughout the entire depicted evolution. The almost perfect periodic oscillations that we find in this case indicate that the linear response term (\ref{expandlin}) is heavily dominated by the contribution of the lowest energy $CT=-1$ state. This is the single-particle vector eigenstate, with $E_1-E_0=M_{v_1}=1.042068$ from our calculations of the excitation energies. This leads to a $\propto (\cos M_{v_1}t-1)$ behavior of the term (\ref{expandlin}), which is indeed what we see in fig.2b.

In fig.2c in the main text we also display the time-evolution of the $CT=+1$ electric field squared operator, $O=\f{1}{2|\mathbb{Z}|}\sum_n L^2(n)$. In that case, the linear response term vanishes entirely because of the $CT$-invariance of $O$. The small $\alpha t$ behavior is then dominated by the $\alpha^2$ term in (\ref{dyson}).

\end{document}